\PassOptionsToPackage{colorlinks=true, citecolor=hotpink, linkcolor=ACMPurple, urlcolor=ACMDarkBlue}{hyperref}
\documentclass[acmsmall,natbib=false]{acmart}
\pdfoutput=1  % for arxiv.org

\setcopyright{acmlicensed}
\copyrightyear{2026}
\acmYear{2026}
\acmDOI{XXXXXXX.XXXXXXX}

\acmVolume{XXX}
\acmNumber{YYY}
\acmArticle{111}
\acmMonth{1}

\newtheorem{theorem}{Theorem}[section]

\newtheorem{corollary}[theorem]{Corollary}
\newtheorem{proposition}[theorem]{Proposition}

\newtheorem*{theoremwithoutno}{Theorem}

\theoremstyle{definition}
\newtheorem{definition}{Definition}[section]

\theoremstyle{remark}
\newtheorem{remark}{Remark}[section]

\newtheorem{theoremsection}{Theorem}[section]

\newtheorem{lemma}[theoremsection]{Lemma}

\def\QEDopen{{\setlength{\fboxsep}{0pt}\setlength{\fboxrule}{0.2pt}\fbox{\rule[0pt]{0pt}{1.3ex}\rule[0pt]{1.3ex}{0pt}}}}
\def\QED{\QEDopen}

\def\Q.E.D{\hfill\QED}

\begin{document} 

\title{Probabilistic Computers (and Hence Quantum Computers) Are Rigorously More Powerful Than Classical Deterministic Computers, and Derandomization}%
%\titlenote{A conference version of this paper appeared in SODA 2021~\cite{mathieu2021competitive}.
%The journal version was published in ACM Transactions on Algorithms,
%\url{https://doi.org/10.1145/3672614} .}

 \date{\today}

\author{Tianrong Lin}
\affiliation{%
 \institution{Hakka University}
 %\city{}
 %\state{}
 \country{China}
}
\orcid{0000-0002-1187-2395}
%\email{}

% \renewcommand{\shortauthors}{Mathieu, Rajaraman, Young, Yousefi}

% \maketitle

\begin{abstract}
 In this paper, we extend the techniques developed in our previous work to construct a probabilistic Turing machine that runs within time $O(n^k)$ for every $k\in\mathbb{N}_1$ and accepts a language $L_d\notin\mathcal{P}$. We further show that $L_d\in\mathcal{BPP}$, thereby separating $\mathcal{BPP}$ from $\mathcal{P}$ (i.e., $\mathcal{P}\subsetneqq\mathcal{BPP}$). 

Since the complexity class $\mathcal{BQP}$ of {\em bounded error quantum polynomial-time computation} contains $\mathcal{BPP}$ (i.e., $\mathcal{BPP}\subseteq\mathcal{BQP}$), our result confirms the long-standing conjecture that quantum computers are {\em rigorously more powerful} than classical deterministic computers (i.e., $\mathcal{P}\subsetneqq\mathcal{BQP}$).

As an important consequence of the above results, we disprove the {\bf Extended Church-Turing Thesis}.

Furthermore, we establish the following separations:
\begin{enumerate}
  \item [(1)]{$\mathcal{P}\subsetneqq\mathcal{RP}$;}
  \item [(2)]{$\mathcal{P}\subsetneqq{\rm co}\mathcal{RP}$;}
  \item [(3)]{$\mathcal{P}\subsetneqq\mathcal{ZPP}$.}
\end{enumerate}

These relationships were long-standing open questions in complexity theory.

In addition, the separation $\mathcal{P}\subsetneqq\mathcal{BPP}$ demonstrates that {\em randomness} plays an essential role in probabilistic computation. In particular, we prove the following:
\begin{enumerate}
  \item [(4)]{The number of random bits used by any probabilistic algorithm accepting $L_d$ cannot be reduced to $O(\log n)$;}
  \item [(5)]{There exists no efficient (complexity-theoretic) {\em pseudorandom generator} (PRG): 
$$
G:\{0,1\}^{O(\log n)}\rightarrow \{0,1\}^n;$$
}
  \item [(6)]{There exists no quick HSG $H:k(n)\rightarrow n$ with $k(n)=O(\log n)$. }
\end{enumerate}
\end{abstract}


%%
%% Keywords. The author(s) should pick words that accurately describe
%% the work being presented. Separate the keywords with commas.

\received{19 July 2026}
\received[revised]{19 July 2026}
\received[accepted]{19 July 2026}

\setcopyright{rightsretained}
%\acmJournal{TOCT}
\acmYear{2026} \acmVolume{1} \acmNumber{1} \acmArticle{1} \acmMonth{1}\acmDOI{10.1145/xxxxxx}

%%
%% This command processes the author and affiliation and title
%% information and builds the first part of the formatted document.
\maketitle
\vskip 0.3 cm
\section{Introduction}
\label{sec:introduction}
\vskip 0.3 cm

A celebrated theorem in complexity theory claims that the existence of pseudorandom generators (PRGs) is equivalent to strong circuit lower bounds (see, e.g., \cite{CT23,CRTY23}):

\begin{theoremwithoutno}[hardness vs. randomness \cite{IW97}]
There exists a Pseudorandom Generator for linear sized circuits that has seed length $l(n)=O(\log n)$, error $1/{\rm poly}(n)$ and running time ${\rm poly}(n)$ if and only if there exists some constant $\epsilon>0$ and  $L\in{\rm DTIME}[2^n]$ such that no circuit of size $2^{\epsilon\cdot n}$ can compute $L$ correctly on infinitely many input lengths.
\end{theoremwithoutno}

Therefore, belief in certain strong circuit lower bounds implies the existence of PRGs sufficient to establish $\mathcal{P}=\mathcal{BPP}$. The $\mathcal{BPP}=\mathcal{P}$ conjecture is, of course, one of the most important open problems in complexity theory (see, e.g., \cite{CT23,CRTY23}) and has driven progress in the field of complexity theory for many decades \cite{CT23}. 

Furthermore, to the best of our knowledge, two fundamental problems in theoretical computer science are derandomization (which concerns the power of probabilistic algorithms) and circuit lower bounds (which concern the power of non-uniform circuits); see, e.g., \cite{CRTY23}. The hardness vs. randomness theorem stated above closely links these two problems.

It is well-known that the class of $\mathcal{BPP}$ (see the popular introduction \cite{Ano4}), originally defined in \cite{Gil77}, consists of the decision problems solvable by probabilistic Turing machines in polynomial time with error probability bounded by $1/3$ for all inputs. In fact, $\mathcal{BPP}$ is one of the largest practical classes of problems: most problems of interest in $\mathcal{BPP}$ have efficient probabilistic algorithms that can be run quickly on probabilistic machines. Moreover, $\mathcal{BPP}$ contains $\mathcal{P}$, the class of problems solvable in polynomial time by deterministic machines, since every deterministic machine is a special case of a probabilistic one (see \cite{Ano2}). However, there are many problems known to lie in $\mathcal{BPP}$ but not known to lie in $\mathcal{P}$.

It is also worth noting that quantum computation has recently received a great deal of attention (see \cite{BV97,BBBV97,FG99,For03,Gro96,Sho97}). A natural question arises: How powerful can {\em quantum computers} be? It is widely believed that quantum computers are at least as powerful as classical computers, owing to the discovery of polynomial-time quantum algorithms for {\em prime factorization} by Shor \cite{Sho97} (see the popular introduction \cite{Ano3}) and $O(\sqrt{N})$ quantum algorithms for {\em database search} by Grover \cite{Gro96}. However, there is still no formal proof that quantum computers are strictly more powerful than our classical computers, because it remains unknown whether prime factorization admits a classical polynomial-time algorithm. Shor's algorithm shows only that prime factorization lies in complexity class $\mathcal{BQP}$ (see the popular introduction \cite{Ano5}), defined in \cite{BV97} as the class of problems solvable in polynomial time by quantum Turing machines with error probability at most $1/3$. Grover's $O(\sqrt{N})$ algorithm, being oracle-based \cite{Gro96,BBBV97}, does not place the problem in $\mathcal{BQP}$. In short, current evidence is insufficient to conclude that quantum computers are rigorously more powerful than classical computers. In other words, it is still unknown whether {\em prime factorization} lies in $\mathcal{P}$.

What, then, is the true relationship between $\mathcal{BPP}$ and $\mathcal{BQP}$? It was proved in \cite{BV97} that $\mathcal{BPP}\subseteq\mathcal{BQP}$. This shows that every language decidable in polynomial time by a probabilistic Turing machine (with error probability bounded at most $1/3$) is also decidable in polynomial time by a quantum Turing machine (with the same error bound), but it says nothing about whether the containment is strict. It has been conjectured \cite{BBBV97} that proving $\mathcal{BQP}\ne\mathcal{BPP}$ would require resolving the major open question $\mathcal{P}\overset{?}{=}\mathcal{PSPACE}$. Indeed, if $\mathcal{P}=\mathcal{PSPACE}$, then $\mathcal{BPP}=\mathcal{BQP}$. However, although we have shown $\mathcal{P}\ne\mathcal{PSPACE}$ \cite{Lin21a,Lin21b}, this does not yet allow us to separate $\mathcal{BQP}$ from $\mathcal{BPP}$, as there is no analogous technique for handling probabilistic Turing machines via universal quantum Turing machines.

Despite the above, we can prove that $\mathcal{BQP}\neq\mathcal{P}$ (or more precisely, $\mathcal{P}\subsetneqq\mathcal{BQP}$) by showing  $\mathcal{P}\subset\mathcal{BPP}$ using extensions of the techniques developed in \cite{Lin21a,Lin21b}. This demonstrates that quantum computers are rigorously more powerful than classical deterministic computers. Prior to this result, the precise relationship between $\mathcal{P}$ and $\mathcal{BQP}$ was unknown. Following Shor's result, it is natural to ask how powerful quantum computers truly are. For example, can all problems in $\mathcal{NP}$ be solved efficiently by quantum computers in polynomial time?  Although \cite{BBBV97} showed that, relative to an oracle chosen uniformly at random with probability $1$, $\mathcal{NP}$ cannot be solved on quantum Turing machine in time $o(2^{n/2})$, this does not resolve whether $\mathcal{NP}\not\subset\mathcal{BQP}$, since oracle separations are not necessary and sufficient conditions. Thus, the exact relationship between $\mathcal{BQP}$ and $\mathcal{NP}$ remains unknown.

Prior to our separation of $\mathcal{P}$ from $\mathcal{BPP}$, many problems were known to lie in $\mathcal{BPP}$ but not known to lie in $\mathcal{P}$. In recent decades, however, significant progress has been made in derandomization (see, e.g., \cite{BFN$^+$93,NW94,IW97,IW01,IKW02,MV99,STV01,ACR98}), providing strong evidence that randomness can often be removed from probabilistic computation. It is widely conjectured that $\mathcal{P}=\mathcal{BPP}$ (see \cite{CRT98} for recent advances towards proving $\mathcal{P}=\mathcal{BPP}$). Loosely speaking,  {\em derandomization} seeks to simulate probabilistic algorithms deterministically, with the central goal being to determine whether $\mathcal{BPP}=\mathcal{P}$. A highlights of this line of work is the construction in \cite{IW97}, which implies $\mathcal{P}=\mathcal{BPP}$ under certain hardness assumptions (see \cite{Imp02} for a survey).

A key tool for derandomization is the (complexity-theoretic) {\em pseudorandom generator} (PRG) \cite{NW94}, which expands a short truly random seed into a longer, pseudorandom string. The (complexity-theoretic) pseudorandom generator is a twist on the definition of a {\em cryptographically secure pseudorandom generator}, with the main difference allowing the generator to run in {\em exponential time} \cite{AB09}. {\em Pseudorandom sequences} produced by PRGs are also important for cryptographic applications such as stream ciphers \cite{CDR04}. It was shown in \cite{ACR98} that {\em quick hitting set generators} (HSGs) can replace PRGs for derandomizing {\em two-sided error} probabilistic algorithms.

Prior to this work, the central open question in {\em derandomization} was to prove an unconditional derandomization result for $\mathcal{BPP}$ or for $\mathcal{ZPP}$ (defined later). Given recent results (see the survey \cite{Kab02}), derandomizing $\mathcal{BPP}$ appears quite difficult. It may be easier to derandomize $\mathcal{ZPP}$, as the assumption $\mathcal{ZPP}=\mathcal{P}$ is not known to imply circuit lower bounds. For more details, see the survey \cite{Kab02}. 

\vskip 0.3 cm
\subsection{Our Contributions}
\vskip 0.3 cm

Motivated by the above considerations, this paper resolves the $\mathcal{BPP}=\mathcal{P}$ conjecture in the negative. Although recent research gives strongly suggests that adding randomness does not increase the class of problems solvable in polynomial time (i.e., that $\mathcal{P}=\mathcal{BPP}$), no proof of this conjecture exists. In light of \cite{Imp02}, possibilities concerning the power of randomized algorithms include: 
\begin{enumerate}
  \item [1.]{Randomization always helps for intractable problems, i.e., $\mathcal{EXP}=\mathcal{BPP}$.}
  \item [2.]{The extent to which randomization helps is problem-specific. It can reduce complexity by any amount from not at all to exponentially.}
  \item [3.]{True randomness is never needed, and random choices can always be simulated deterministically, i.e., $\mathcal{P}=\mathcal{BPP}$.}
\end{enumerate} 

We disprove the third possibility with the following result: 

\begin{theorem}
\label{theorem1.1}
 There exists a language $L_d$ that is not accepted by any polynomial-time deterministic Turing machines but is accepted by a probabilistic Turing machine $M_0$ running within time $O(n^k)$ for any $k\in\mathbb{N}_1$ with probability at least $\frac{2}{3}$. Moreover, $L_d\in\mathcal{BPP}$. In other words, 
$$
\mathcal{P}\subsetneqq\mathcal{BPP}.
$$
\end{theorem}
\noindent Combined with the known inclusion $\mathcal{BPP}\subseteq\mathcal{BQP}$ \cite{BV97}, this immediately yields:
\begin{corollary}
\label{corollary1.2}
  $\mathcal{P}\subsetneqq\mathcal{BQP}$.\Q.E.D
\end{corollary}

Thus, both quantum computers and classical probabilistic computers are rigorously more powerful than deterministic classical computers.

Another interesting consequence concerns the Extended Church-Turing Thesis (see, e.g., p. 23 in \cite{DK14}):

\indent{\bf Extended Church-Turing Thesis.} A function computable in polynomial time in any {\em reasonable} computational model using a {\em reasonable} time complexity measure is computable by a deterministic Turing machine in polynomial time.

The intuitive notion of {\em reasonable computational models} excludes any machine models that are not realizable by physical devices. In particular, we do not consider a nondeterministic Turing machine to be a reasonable model. We also exclude any time complexity measures that do not reflect the physical time requirements (these mean that proving $\mathcal{P}\neq\mathcal{NP}$ \cite{Lin21a, Lin24} does not disprove the {\em Extended Church-Turing Thesis}). However, since quantum Turing machines \cite{BV97} are considered physically realizable (see, e.g., \cite{Yao93}) and we have shown $\mathcal{P}\subsetneqq\mathcal{BQP}$, the {\em Extended Church-Turing Thesis} is indeed false.

\begin{corollary}
The {\bf Extended Church-Turing Thesis} is false.
\end{corollary}

We next consider the class $\mathcal{RP}$, defined as the set of languages $L$ for which there exists a probabilistic polynomial-time Turing machine $A$ such that 

$$
x\in L\Rightarrow\mathrm{Pr}[A(x) \text{ accepts}]\geq\frac{1}{2}
$$
and
$$
x\not\in L\Rightarrow\mathrm{Pr}[A(x) \text{ accepts}]=0.
$$

Using similar techniques, we obtain:

\begin{theorem}
\label{theorem1.3}
  There exists a language $\widetilde{L_d}$ that is not accepted by any polynomial-time deterministic Turing machine  but is accepted by a probabilistic Turing machine $M'_0$ running within time $O(n^k)$ for any $k\in\mathbb{N}_1$ with probability at least $\frac{1}{2}$. Moreover, $\widetilde{L_d}\in\mathcal{RP}$.
\end{theorem}

 This immediately implies:
\begin{corollary}
\label{corollary1.4}
$\mathcal{P}\subsetneqq\mathcal{RP}$.\Q.E.D
\end{corollary}

Recall that the complement of the complexity class $\mathcal{RP}$ (the definition of the complement of a complexity class is given in subsection \ref{subsec:com_complexity_class}), denoted ${\rm co}\mathcal{RP}$, consists of all languages $L$ for which there exists a probabilistic polynomial-time Turing machine $A$ such that
$$
x\in L\Rightarrow\mathrm{Pr}[A(x) \text{ accepts}]=1
$$
and
$$
x\not\in L\Rightarrow\mathrm{Pr}[A(x) \text{ accepts}]\leq \frac{1}{2}.
$$
Interestingly, we show that slightly modified arguments from the proofs of Theorem \ref{theorem1.1} and Theorem \ref{theorem1.3} can also be used to establish $\mathcal{P}\subsetneqq{\rm co}\mathcal{RP}$. More precisely: 

\begin{theorem}
\label{theorem1.5}
  There exists a language $\widehat{L_d}$ that is not accepted by any polynomial-time deterministic Turing machine but is accepted by a probabilistic Turing machine $N'_0$ running within time $O(n^k)$ for any $k\in\mathbb{N}_1$ with probability $1$. Moreover, $\widehat{L_d}\in{\rm co}\mathcal{RP}$.
\end{theorem}

This yields:
\begin{corollary}
\label{corollary1.6}
$\mathcal{P}\subsetneqq{\rm co}\mathcal{RP}$.\Q.E.D
\end{corollary}

Interestingly, the language $\widehat{L_d}$ satisfies the following properties (see Remark \ref{remark7.1}):

For $x\in\widehat{L_d}$, 
$$
{\rm Pr}[N'_0(x)\text{ accepts}]=1\geq\frac{1}{2};
$$
for $x\not\in\widehat{L_d}$,
$$
{\rm Pr}[N'_0(x)\text{ accepts}]=0.
$$
This shows that $\widehat{L_d}$ is also in $\mathcal{RP}$. Denoting $\mathcal{RP}\cap{\rm co}\mathcal{RP}$ by $\mathcal{ZPP}$, we obtain a separation of $\mathcal{ZPP}$ from  $\mathcal{P}$.

\begin{corollary}
\label{corollary1.7}
$\mathcal{P}\subsetneqq\mathcal{ZPP}$. \Q.E.D
\end{corollary}

Generally, {\em enumeration}\footnote{Not to be confused with the {\em enumeration} discussed in Section \ref{sec:enumeration}, which refers to establish a $(1,1)$ correspondence between $\mathbb{N}_1$ and the set $\{(M,k)\}$ of all polynomial-time deterministic Turing machines.} (see p. 51 in \cite{Vad12}) is a derandomization technique that allows us to deterministically simulate any randomized algorithm with an exponential slowdown: 
$$
\mathcal{BPP}\subseteq\mathcal{EXP}\overset{\rm def}{=}\bigcup_{c\in\mathbb{N}_1}{\rm DTIME}[2^{n^c}].
$$
This approach is general in that it applies to all $\mathcal{BPP}$ algorithms. However, if an algorithm for a language $$L\in\mathcal{BPP}$$ uses only a small number of random bits (say $O(\log n)$), then $L\in\mathcal{P}$ (see Proposition 3.3, p. 52 in \cite{Vad12}). Take the language $L_d$ from Theorem \ref{theorem1.1} as an example; we prove the following result, which demonstrates that {\it randomness} is an important computational resource:

\begin{theorem}
\label{theorem1.8}
No probabilistic algorithm accepting $L_d$ can use only $O(\log n)$ random bits.
\end{theorem}

Impagliazzo and Wigderson \cite{IW97} showed that if
$$
{\rm E}\overset{\rm def} {=}{\rm DTIME} [2^{O(n)}]
$$
has a function of circuit complexity $2^{\Omega(n)}$, then
$$
\mathcal{BPP}=\mathcal{P}.
$$

Theorem \ref{theorem1.1} immediately implies the following

\begin{corollary}
\label{corollary1.9}
No function in $\text{\em E}$ has circuit complexity $2^{\Omega(n)}$.\Q.E.D
\end{corollary}

As we mentioned earlier, a main basic method used to derandomize algorithms is to use a (complexity-theoretic) {\em pseudorandom generator} (PRG), and if there exists some specific PRGs, then
$$
\mathcal{BPP}=\mathcal{P}.
$$

We show the following:

\begin{theorem}
\label{theorem1.10}
There exists no (complexity-theoretic) pseudorandom generator
$$
G:\{0,1\}^{l(t)}\rightarrow\{0,1\}^t
$$
with
$$
l(t)=O(\log t).
$$
\end{theorem}

In \cite{ACR98}, it showed that {\em quick hitting set generators} (HSGs) can replace {\em quick pseudorandom generators} to derandomize any probabilistic {\em two-sided error} algorithms. An important result in \cite{ACR98} is that if a logarithmic price quick-hitting set generator exists, then $\mathcal{BPP}=\mathcal{P}$. By these, we can show such an HSG does not exist.

\begin{theorem}
\label{theorem1.11}
 Let
 $$
 k(n)=O(\log n).
 $$
 Then there exists no quick HSG
 $$
 H:k(n)\rightarrow n.
 $$
\end{theorem}

\vskip 0.3 cm
\subsection{Organization}
\label{subsec:overview}
\vskip 0.3 cm

The rest of the paper is organized as follows. In Section \ref{sec:definitions_and_notation} we review relevant definitions, fix notation, and present several technical lemmas. In Section \ref{sec:enumeration} we establish a bijection between $\mathbb{N}_1$ and the set of all polynomial-time deterministic Turing machines. Section \ref{sec:diagonalization} contains the construction of the probabilistic Turing machine that accepts a language $L_d\notin\mathcal{P}$. In Section \ref{sec:l_dinbpp} we prove that $L_d\in\mathcal{BPP}$. Theorems \ref{theorem1.3} and \ref{theorem1.5} are proved in Sections \ref{sec:proof_of_pneqrp} and \ref{sec:proof_of_pneqcorp}, respectively. The proofs of Theorems \ref{theorem1.8}, \ref{theorem1.10}, and \ref{theorem1.11} appear in Section \ref{sec:randomness_helps}. We conclude in the final section.

\vskip 0.3 cm
\section{Preliminaries}
\label{sec:definitions_and_notation}
\vskip 0.3 cm

In this section, we describe the notation and notions needed in the following.

Let $\mathbb{N}$ denote the natural numbers $$\{0,1,2,3,\cdots\}$$ where $+\infty\not\in\mathbb{N}$. Furthermore, $\mathbb{N}_1$ denotes the set  $$\mathbb{N}\setminus\{0\}, $$ i.e., the set $$\{1,2,3,\cdots\}.$$ It is clear that there is a bijection between $\mathbb{N}$ and $\mathbb{N}_1$. To see this, it suffices to take the bijection $$n\mapsto n+1,$$ where $n\in\mathbb{N}$ and $n+1\in\mathbb{N}_1$. 

The big $O$ notation describes the order of growth of some quantity as a function of $n$ or the limiting behavior of a function. For example, $S(n)$ is big $O$ of $f(n)$, written, $$S(n)=O(f(n)), $$ if there exists a positive integer $N_0$ and a positive constant $M$ such that $$S(n)\leq M\times f(n)$$ for all $n>N_0$. 

The little $o$ notation also describes the order of growth of some quantity as a function of $n$ or the limiting behavior of a function, but with different meaning. Specifically, $T(n)$ is little $o$ of $t(n)$, written $$T(n)=o(t(n))$$ if for any constant $c>0$, there exists a positive integer $N_0>0$ such that $$T(n)<c\times t(n)$$ for all $n>N_0$.

The big $\Omega$ notation likewise describes the limiting behavior of a function of $n$ with different means. Specifically, $t(n)$ is big $\Omega$ of $g(n)$, written, $$t(n)\in\Omega(g(n)) $$ if there exists a positive integer $N_0$ and a positive constant $c$ such that $$t(n)>c\times g(n)$$ for all $n>N_0$.

The computation models we use  are the {\em Turing machine} as  defined in standard textbooks such as \cite{AHU74}, the {\em quantum Turing machines} as  defined in \cite{BV97}, and the {\em probabilistic Turing machines} as  defined in \cite{San69,San71,Gil77}.

\vskip 0.3 cm
\subsection{Polynomial-time Deterministic Turing Machines}
\label{subsec:pdtms}
\vskip 0.3 cm

To define polynomial-time deterministic Turing machines, we first recall the definition of a deterministic Turing machine:

\begin{definition}[$k$-tape deterministic Turing machine, \cite{AHU74}]
\label{definition2.1}
 A $k$-tape deterministic Turing machine (shortly, DTM) $M$ is a seven-tuple $$(Q,T,I,\delta,\mathrm{b},q_0,q_f)$$
where :
 \begin{enumerate}
 \item {$Q$ is the set of states.}
 \item {$T$ is the set of tape symbols.}
 \item {$I$ is the set of input symbols; $I\subseteq T$.}
 \item {$\mathrm{b}\in T-I$ is the blank.}
 \item {$q_0$ is the initial state.}
 \item {$q_f$ is the final (or accepting) state.}
 \item {$\delta$ is the next-move function, maps a subset of $Q\times T^k$ to
 $$
 Q\times(T\times\{L,R,S\})^k.
 $$
  That is, for some $(k+1)$-tuples consisting of a state and $k$ tape symbols, it gives a new state and $k$ pairs, each pair consisting of a new tape symbol and a direction for the tape head. Suppose $$\delta(q,a_1,a_2,\cdots,a_k)=(q',(a'_1,d_1),(a'_2,d_2),\cdots,(a'_k,d_k)),$$
 and the deterministic Turing machine is in state $q$ with the $i$th tape head scanning tape symbol $a_i$ for $1\leq i\leq k$. Then in one move the deterministic Turing machine enters state $q'$, changes symbol $a_i$ to $a'_i$, and moves the $i$th tape head in the direction $d_i$ for $1\leq i\leq k$.}
  \end{enumerate}
 \end{definition}

 \vskip 0.3 cm

 The notion of a nondeterministic Turing machine is similar to that of a deterministic Turing machine, except that the next-move function $\delta$ is a mapping from $$Q\times T^k$$ to subsets of $$ Q\times(T\times\{L,R,S\})^k, $$ stated as follows:

 \begin{definition}[$k$-tape nondeterministic Turing machine, \cite{AHU74}]
 \label{definition2.2}
A $k$-tape nondeterministic Turing machine (shortly, NTM) $M$ is a seven-tuple $$(Q,T,I,\delta,\mathrm{b},q_0,q_f)$$
where all components have the same meaning as for the ordinary deterministic Turing machine, except that here the next-move function $\delta$ is a mapping from $$Q\times T^k$$ to subsets of $$Q\times(T\times\{L,R,S\})^k.$$
\end{definition}

In the remainder of this paper, ``Turing machine" refers to a deterministic Turing machine unless otherwise specified, and we sometimes use DTM as an abbreviation.

Now, we give the definition of polynomial-time deterministic Turing machines as follows:

\begin{definition}[cf. \cite{Coo00}]
\label{definition2.3}
Formally, a polynomial-time deterministic Turing machine is a deterministic Turing machine $M$ such that there exists $k\in\mathbb{N}_1$, for all input $x$ of length $n$ where $n\in\mathbb{N}$ is arbitrary, $M(x)$ will halt within $n^k+k$ steps.
\end{definition}

We represent such a machine by a pair $(M,k)$, where $M$ is the polynomial-time deterministic Turing machine itself and $k$ is the smallest integer such that $M$ runs within time $n^k+k$ on all inputs of length $n$. We call $k$ the order of the polynomial-time deterministic Turing machine $(M,k)$.

The family of languages decidable in deterministic time $T(n)$ is denoted by ${\rm DTIME}[T(n)]$.

\subsection{Quantum Turing Machines}
\vskip 0.3 cm

We now recall the definition of a {\em quantum Turing machine}:

\begin{definition}[cf. Definition 3.2 in \cite{BV97}]
\label{definition2.4}
Let $\widetilde{\mathbb{C}}$ be the set consisting of $\alpha\in\mathbb{C}$ such that there is a deterministic algorithm that computes the real and imaginary parts of $\alpha$ to within $2^{-n}$ in time polynomial in $n$. A QTM $M$ is defined by a triplet $(I,\Gamma,Q,\delta)$, where $I\subset\Gamma$ is the finite input alphabet and $\mathrm{b}\not\in I$, $\Gamma$ is the finite tape alphabet and $\mathrm{b}\in\Gamma$ is the blank symbol, $Q$ is a finite set of states with an identified initial state $q_0$ and final state $q_f\neq q_0$, and $\delta$, the quantum transition function, is a function $$\delta: Q\times\Gamma\rightarrow\widetilde{\mathbb{C}}^{\Gamma\times Q\times\{L,R,S\}}. $$

The QTM has a two-way infinite tape of cells indexed by $\mathbb{Z}$ and a single read/write tape head that moves along the tape.

Let $S$ be the inner-product space of finite complex linear combinations of configurations of $M$ with the Euclidean norm. We call each element $\psi\in S$ a superposition of $M$. Then QTM $M$ defines a linear operator $$U_M:S\rightarrow S,$$ called the time evolution operator of $M$, as follows: If $M$ starts in configuration $c$ with current state $p$ and scanned symbol $\sigma$, then after one step $M$ will be in superposition of configurations $$\phi=\sum_i\alpha_ic_i,$$ where each nonzero $\alpha_i$ corresponds to a transition $\delta(p,\sigma,\tau,q,d)$, and $c_i$ is the new configuration that results from applying this transition to $c$. Extending this map to the entire space $S$ through linearity gives the linear time evolution operator $U_M$ such that $U_M$ is unitary.
\end{definition}

The class of languages ${\rm BQTIME}[T(n)]$ decided by quantum Turing machines with bounded error within time $T(n)$ is defined as follows:

\begin{definition}
\label{definition2.5}
 A language $L$ is in ${\rm BQTIME}[T(n)]$ if and only if there exists a quantum Turing machine $M$, such that
 \begin{itemize}
  \item [1.]{ $M$ runs in time $T(n)$ on all inputs.}
  \item [2.]{ For all $x\in L$, $M$ accepts $x$ with probability $\geq\frac{2}{3}$.}
  \item [3.]{ For all $x\not\in L$, $M$ accepts $x$ with probability $\leq\frac{1}{3}$.}
 \end{itemize}
 \end{definition}

\subsection{Probabilistic Turing Machines}
\vskip 0.3 cm

A {\em probabilistic Turing machine} is a nondeterministic Turing machine that chooses among the available transitions at each step according to a probability distribution (see \cite{Ano2} for a popular introduction).

There are several equivalent definitions of probabilistic Turing machines in the literature, such as the one in \cite{DK14}, the one presented in \cite{San69, San71}, and the one given in \cite{Gil77}. Note that the definition given in \cite{San71} is more general than that given in \cite{Gil77}. We primarily follow the formulation in \cite{San71} and later show its equivalence to the common random-tape model in Appendix \ref{sec:appendix}. 

\begin{definition}[\cite{San71}, Definition 3.1]
\label{definition2.6}
A probabilistic Turing machine (PTM) may be defined through the specification of three mutually disjoint finite nonempty sets $A$, $B$, and $S$; a function $p$ from $S\times U\times V\times S$ into $[0,1]$, where $U=A\cup B$, $V=U\cup S\cup\{+,-,\cdot\}$,$+$, $-$, $\cdot\not\in U\cup S$; and a function $h$ from $S$ into $[0,1]$. The functions $p$ and $h$ satisfy the following conditions:
\begin{itemize}
\item[1.] {$\sum_{v\in V}\sum_{s'\in S}p(s,u,v,s')=1$ for every $s\in S$, $u\in U$, and }
\item[2.] {$\sum_{s\in S}h(s)=1$.}
\end{itemize}
The sets $A$ and $B$ are, respectively, the printing and auxiliary alphabets. The set $S$ is the set of internal states. $h(s)$ is the probability that the initial state is $s$, and $p(s,u,v,s')$ gives the probability of the ``next act" of the PTM given that its present state is $s$ and input $u$ is applied. The ``next act" of a PTM is determined by $v$ and may be any one of the conventional Turing machine operations.
\begin{itemize}
\item [1.]{$v\in U$: replace $u$ by $v$ on the scanned square and go to state $s'$.}
\item [2.]{$v=+$: move one square to the right and go to state $s'$.}
\item [3.]{$v=-$: move one square to the left and go to state $s'$.}
\item [4.]{$v=\cdot$: stop.}
\item [5.]{$v\in S$: go to either $v$ or $s'$ depending on a given random set.}
\end{itemize}
The functions $p$ and $h$ will be referred to as the transition function and initial distribution, respectively. If $h$ is concentrated at a single state $s_0\in S$, i.e., $h(s_0)=1$ and $h(s)=0$ for $s\neq s_0$, then we say that $s_0$ is the initial state.
\end{definition}

\begin{definition}[\cite{San71}, Definition 3.2]
   \label{definition2.7}
   Let $Z=(A,B,S,p,h)$ be a PTM. Then
   \begin{itemize}
     \item [1.]{$Z$ is deterministic iff the range of both $p$ and $h$ consists of only two numbers, $0$ and $1$.}
     \item [2.]{$Z$ is simple iff $p(s,u,v,s')=0$ for every $s,s'\in S$, $u\in A\cup B$, and $v\in S$.}
   \end{itemize}
   \end{definition}

 \begin{remark}
 \label{remark1}
 Observe that the conventional Turing machines are deterministic PTM. In the case of a deterministic PTM, the transition function $p$ is uniquely determined by the set $$\{(s,u,v,s')\text{ : } p(s,u,v,s')=1\,\,{\rm and}\,\,v\neq\cdot\}.$$
\end{remark}
\vskip 0.3 cm

By Definition \ref{definition2.6} and Definition \ref{definition2.7}, we can adapt the definition of a {\em probabilistic Turing machine} from the nondeterministic Turing machine given in Definition \ref{definition2.2} as follows:

\begin{definition}[Probabilistic Turing Machine (adaptation from Definition \ref{definition2.6})]
\label{definition2.8}
A $k$-tape probabilistic Turing machine (shortly, PTM) $M$ is an $8$-tuple. $$(Q,T,I,\mathrm{b},q_0,q_f,\delta),$$
where:
   \begin{enumerate}
   \item {$Q$ is the set of states.}
   \item {$T$ is the set of tape symbols.}
   \item {$I$ is the set of input symbols; $I\subseteq T$.}
   \item {$\mathrm{b}\in T-I$ is the blank.}
   \item {$q_0$ is the initial state.}
   \item {$q_f$ is the final (or accepting) state.}
   \item {$\delta$ is the probabilistic transition, defined as: $$Q\times T^k\to\mathcal{P}(Q\times(T\times\{L,R,S\})^k),$$ where $\mathcal{P}(S)$ denotes the set of probability distribution over $$S=Q\times(T\times\{L,R,S\})^k.$$ For each
       $$(q,a_1,a_2,\cdots,a_k)\in Q\times T^k,$$ $\delta(q,a_1,a_2,\cdots,a_k)$ assigns to every possible transition 
       $$
       (q,a_1,a_2,\cdots,a_k)\rightarrow(q'_i,(a'_{1,i},d_{1,i}),(a'_{2,i},d_{2,i}),\cdots,(a'_{k,i},d_{k,i})),
       $$ a probability $$p((q,a_1,a_2,\cdots,a_k),(q'_i,(a'_{1,i},d_{1,i}),(a'_{2,i},d_{2,i}),\cdots,(a'_{k,i},d_{k,i})))$$ such that
       $$
       \sum_{(q'_i,(a'_{1,i},d_{1,i}),(a'_{2,i},d_{2,i}),\cdots,(a'_{k,i},d_{k,i}))} p((q,a_1,a_2,\cdots,a_k),(q'_i,(a'_{1,i},d_{1,i}),(a'_{2,i},d_{2,i}),\cdots,(a'_{k,i},d_{k,i})))=1,
       $$
       i.e., the sum is taken over all possible transitions.
       }
\end{enumerate}
\end{definition}

Setting $k=1$ in Definition \ref{definition2.8} yields the definition of single-tape probabilistic Turing machines.

Another common definition of probabilistic Turing machines, widely used in randomized computation, is the following random-tape model:

\begin{definition}[\cite{DK14}, Section 8.2]
\label{definition2.ten}
A multi-tape {\it probabilistic Turing machine} (PTM) $M$ consists of two components $(\widetilde{M},\phi)$, where $\widetilde{M}$ is a regular multi-tape deterministic Turing machine and $\phi$ is a random-bit generator. In addition to the input, output, and work tapes, the machine $M$ has a special {\it random-bit} tape. The computation of the machine $M$ on an input $x$ can be informally described as follows: At each move, the generator $\phi$ first writes a random bit $b=0$ or $1$ on the square currently scanned by the tape head of the random-bit tape. Then the DTM $\widetilde{M}$ makes the next move according to the current state and the current symbols scanned by the tape heads of all (but output) tapes, including the random-bit tape. The tape head of the random-bit tape always moves to the right after each move.
\end{definition}

We will prove that Definition \ref{definition2.8} (probabilistic transitions) and Definition \ref{definition2.ten} (random-tape model) are equivalent in the Appendix \ref{sec:appendix}. This equivalence allows us to use whichever formulation is more convenient when constructing probabilistic Turing machines for our proofs.

The family of languages decidable by probabilistic Turing machines with bounded error (BPTIME) is defined as follows:

\begin{definition}
 \label{definition2.9}
 A language $L$ belongs to ${\rm BPTIME}[T(n)] $ if and only if there exists a probabilistic Turing machine $M$ such that
   \begin{itemize}
      \item [1.]{ $M$ runs within time $T(n)$ on all inputs.}
      \item [2.]{ For all $x\in L$, $M$ accepts $x$ with probability $\geq\frac{2}{3}$.}
      \item [3.]{ For all $x\not\in L$, $M$ accepts $x$ with probability $\leq\frac{1}{3}$.}
   \end{itemize}
\end{definition}

\subsection{Complexity classes $\mathcal{P}$, $\mathcal{BPP}$, and $\mathcal{BQP}$}
\vskip 0.3 cm

Let $w$ be an input string and $|w|$ its length. If, for every input $w$ of length $n$, all computations of a deterministic (respectively, probabilistic or quantum) Turing machine $M$ halt in at most $T(n)$ steps, then $M$ is said to be a $T(n)$ time-bounded deterministic (respectively, probabilistic or quantum) Turing machine, or to have time complexity {\em of time complexity $T(n)$.}

The complexity classes $\mathcal{P}$ and $\mathcal{BPP}$ (see \cite{Ano4} for a popular introduction) are defined as: $$\mathcal{P}=\bigcup_{k\in\mathbb{N}_1}{\rm DTIME}[n^k],$$ and $$\mathcal{BPP}=\bigcup_{k\in\mathbb{N}_1}{\rm BPTIME}[n^k]. $$

Similarly, $$\mathcal{BQP}=\bigcup_{k\in\mathbb{N}_1}{\rm BQTIME}[n^k]. $$

 \subsection{Complement of a Complexity Class}
 \label{subsec:com_complexity_class}
 \vskip 0.3 cm
 
 For a complexity class $\mathcal{C}$, its complement is denoted by ${\rm co}\mathcal{C}$ (see \cite{Pap94}), i.e., $${\rm co}\mathcal{C}=\{\overline{L}:L\in\mathcal{C}\},$$ where $L$ is a decision problem, and $\overline{L}$ is the complement of $L$. For example, ${\rm co}\mathcal{P}$ is the complement of $\mathcal{P}$ , and ${\rm co}\mathcal{NP}$ is the complement of $\mathcal{NP}$. Note that the complement of a decision problem $L$ is defined as the decision problem whose answer is ``{\em yes}" whenever the input is a ``{\em no}" input of $L$ and vice versa.

 \subsection{Complexity Classes $\mathcal{RP}$, ${\rm co}\mathcal{RP}$ and $\mathcal{ZPP}$}
 \vskip 0.3 cm

With the above definitions, the complexity classes of $\mathcal{RP}$, ${\rm co}\mathcal{RP}$, and $\mathcal{ZPP}$ can be defined as follows:
\begin{definition}[cf. Definition 2.9 in \cite{Vad12}]
 \label{definition2.10}
 $\mathcal{RP}$ is the class of languages $L$ for which there exists a polynomial-time probabilistic Turing machine $M$ such that
 \begin{itemize}
 \item [1.]{ $M$ runs in polynomial time in the input size on all inputs.}
 \item [2.]{$x\in L$ if $M$ accepts $x$ with probability $\geq\frac{1}{2}$.}
 \item [3.]{$x\not\in L$ if $M$ accepts $x$ with probability $0$.}
 \end{itemize}
 \end{definition}
 
\begin{definition}
 \label{definition2.t12}
 A language $L$ belongs to ${\rm RPTIME}[T(n)] $ if and only if there exists a probabilistic Turing machine $M$ such that
   \begin{itemize}
      \item [1.]{ $M$ runs within time $T(n)$ on all inputs.}
      \item [2.]{ For all $x\in L$, $M$ accepts $x$ with probability $\geq\frac{1}{2}$.}
      \item [3.]{ For all $x\not\in L$, $M$ accepts $x$ with probability $\leq 0$.}
   \end{itemize}
\end{definition}

By Definition \ref{definition2.t12}, the class $\mathcal{RP}$ is defined as
$$
\mathcal{RP}=\bigcup_{k\in\mathbb{N}_1}{\rm RPTIME}[n^k].
$$This is consistent with Definition \ref{definition2.10}.

Then class ${\rm co}\mathcal{RP}$ is the complement of $\mathcal{RP}$. A more intuitive definition is the following:
\begin{definition}[cf. Definition 2.11 in \cite{Vad12}]
\label{definition2.11}
${\rm co}\mathcal{RP}$ is the class of languages $L$ for which there exists a polynomial-time
probabilistic Turing machine $M$ such that
\begin{itemize}
\item [1.]{ $M$ runs in polynomial time in the input size on all inputs.}
\item [2.]{$x\in L$ if $M$ accepts $x$ with probability $1$.}
\item [3.]{$x\not\in L$ if $M$ accepts $x$ with probability $\leq\frac{1}{2}$.}
\end{itemize}
\end{definition}

\begin{definition}
 \label{definition2.fourty}
 A language $L$ belongs to ${\rm coRPTIME}[T(n)] $ if and only if there exists a probabilistic Turing machine $M$ such that
   \begin{itemize}
      \item [1.]{ $M$ runs within time $T(n)$ on all inputs.}
      \item [2.]{ For all $x\in L$, $M$ accepts $x$ with probability $1$.}
      \item [3.]{ For all $x\not\in L$, $M$ accepts $x$ with probability $\leq \frac{1}{2}$.}
   \end{itemize}
\end{definition}

By Definition \ref{definition2.fourty}, the class ${\rm co}\mathcal{RP}$ is defined as
$$
{\rm co}\mathcal{RP}=\bigcup_{k\in\mathbb{N}_1}{\rm coRPTIME}[n^k].
$$This is consistent with Definition \ref{definition2.11}.

For simplicity, the class $\mathcal{ZPP}$ can equivalently be defined as $\mathcal{RP}\cap{\rm co}\mathcal{RP}$:
\begin{definition}
\label{definition2.12}
$\mathcal{ZPP}\overset{\textit{\rm def}}{=}\mathcal{RP}\cap{\rm co}\mathcal{RP}$.
\end{definition}

\subsection{Useful Lemmas}
\vskip 0.3 cm

Regarding the time complexity of $k$-tape deterministic (or, nondeterministic) Turing machines versus single-tape deterministic (or, nondeterministic) Turing machines, we quote the following useful lemma extracted from \cite{AHU74} (see Lemma 10.1 and Corollary 1 to Lemma 10.1 in \cite{AHU74}), which will play an important role in the sequel:

 \begin{lemma}[Lemma 10.1 in \cite{AHU74}]
 \label{lemma2.1}
 If $L$ is accepted by a $k$-tape nondeterministic $T(n)$ time-bounded Turing machine, then $L$ is accepted by a single-tape nondeterministic $O(T^2(n))$ time-bounded Turing machine.\Q.E.D
 \end{lemma}

 The deterministic version of the above lemma is as follows:

 \begin{corollary}
 \label{corollary2.2}
 If $L$ is accepted by a $k$-tape deterministic $T(n)$ time-bounded Turing machine, then $L$ is accepted by a single-tape deterministic $O(T^2(n))$ time-bounded Turing machine.\Q.E.D
 \end{corollary}

 The above corollary is Corollary 1 in \cite{AHU74} to Lemma 10.1; see also Theorem 6 in \cite{HS65} and Theorem 2.1 in \cite{Pap94}.

 \subsection{Efficient Simulation}
 \vskip 0.3 cm

 The following theorem about efficient simulation is needed a few times, and its proof is present in \cite{HS66}; see also \cite{AB09}.

 \begin{lemma}[\cite{AB09}, see also \cite{HS66}]
 \label{lemma2.3}
 There exists a Turing machine $U$ such that for every $x,\alpha\in\{0,1\}^*$, $U(x,\alpha)=M_{\alpha}(x)$, where $M_{\alpha}$ denotes the Turing machine represented by $\alpha$. Moreover, if $M_{\alpha}$ halts on input $x$ within $T(|x|)$ steps, then $U(x,\alpha)$ halts within $cT(|x|)\log T(|x|)$ steps, where $\log n$ means $\log_2 n$ and $c$ is a constant independent of $|x|$ and depending only on $M_{\alpha}$'s alphabet size, number of tapes, and number of states.\Q.E.D
 \end{lemma}

 Finally, more information and premise lemmas will be given along the way to prove our main result.

\vskip 0.3 cm
\section{Enumeration of Polynomial-Time Deterministic Turing Machines}
\label{sec:enumeration}
\vskip 0.3 cm

In the following context, if a polynomial-time Turing machine (see Definition \ref{definition2.3}) runs in at most $|x|^k$ steps for any input $x$, then we often say that it runs within time $O(n^{k-1})$ rather than $O(n^k)$.

We have represented a polynomial-time deterministic Turing machine by a tuple of $(M,k)$ in subsection \ref{subsec:pdtms}, where $M$ is the polynomial-time deterministic Turing machine itself and $k$ is the smallest integer such that $M(w)$ will run at most $|w|^k+k$ steps for all input $w$.

\vskip 0.3 cm
\begin{remark}
\label{remark3.1}
One convenience of representing polynomial-time deterministic Turing machines by tuples $(M,k)$ is that it allows us to conveniently control the running time of the probabilistic Turing machine $M_0$ constructed in Theorem \ref{theorem4.1} in Section \ref{sec:diagonalization} below. This facilitates our analysis of the time complexity of $M_0$ and, in particular, makes it easier to prove Theorem \ref{theorem4.2}.
\end{remark}
\vskip 0.3 cm

By Corollary \ref{corollary2.2}, we may restrict attention to single-tape deterministic Turing machines. Thus, in what follows, by ``polynomial-time deterministic Turing machines" we mean single-tape polynomial-time deterministic Turing machines.

To obtain our main result, we need to {\em enumerate} all polynomial-time deterministic Turing machines so that for each nonnegative integer $i$ there is a unique tuple $(M,k)$ associated with $i$. In other words, we define a function from $\mathbb{N}_1$ onto the set of all polynomial-time deterministic Turing machines $\{(M,k)\}$) that allows us to refer unambiguously to the $j$-th polynomial-time deterministic Turing machine. For the reader's convenience, we recall the notion of an enumeration:

\begin{definition}[\cite{Rud76}, p. 27, Definition 2.7]
\label{definition3.1}
By an enumeration of set $T$, we mean a function $e$ defined on the set $\mathbb{N}_1$ of all positive integers. If $e(n)=x_n\in T$, for $n\in\mathbb{N}_1$, it is customary to denote the enumeration $e$ by the symbol $\{x_n\}$, or sometimes by $x_1$, $x_2$, $x_3$, $\cdots$. The values of $e$, that is, the elements $x_n\in T$, are called the {\em terms} of the enumeration.
\end{definition}

To achieve this, we first use the encoding method presented in \cite{AHU74}, p. 407, to represent a single-tape deterministic Turing machine into an integer.

Without loss of generality, we make the following assumptions about the representation of a single-tape deterministic Turing machine with input alphabet $\{0,1\}$ because that will be all we need:
\begin{enumerate}
\item {The states are named
$$
q_1,q_2,\cdots,q_s
$$
for some $s$, with $q_1$ the initial state and $q_s$ the accepting state.}
\item {The input alphabet is $\{0,1\}$.}
\item {The tape alphabet is
$$
\{X_1,X_2,\cdots,X_t\}
$$
for some $t$, where $X_1=\mathrm{b}$, $X_2=0$, and $X_3=1$.}
\item {The next-move function $\delta$ is a list of quintuples of the form
$$
(q_i,X_j,q_k,X_l,D_m),
$$
meaning that
$$
\delta(q_i,X_j)=(q_k,X_l,D_m),
$$
and $D_m$ is the direction, $L$, $R$, or $S$, if $m=1,2$, or $3$, respectively. We assume this quintuple is encoded by the string
$$
10^i10^j10^k10^l10^m1.
$$
}
\item {The deterministic Turing machine itself is encoded by concatenating in any order the codes for each of the quintuples in its next-move function. Additional $1$'s may be prefixed to the string if desired. The result will be some string of $0$'s and $1$'s, beginning with $1$, which we can interpret as an integer.}
\end{enumerate}

We further encode the order of $(M,k)$ by the string $$10^k1$$. The full encoding of the tuple $(M,k)$ is then the concatenation of the binary string for $M$ followed by $10^k1$, which itself is interpreted as a positive integer. 

Any integer that cannot be decoded under this scheme is deemed to represent the trivial polynomial-time deterministic Turing machine with an empty next-move function. Every polynomial-time deterministic Turing machine appears infinitely often in the enumeration, since given a polynomial-time deterministic Turing machine, we can prefix $1$'s at will to find larger and larger integers representing the same set of the polynomial-time deterministic Turing machine of $(M,k)$. We denote such a set of the polynomial-time deterministic Turing machine by $\widehat{M}_j$, where $j$ is the integer representing $(M,k)$. The reader will easily get that we have defined a surjective function $e$ from $\mathbb{N}_1$ to the set $\{(M,k)\}$ of all polynomial-time deterministic Turing machines, which is consistent with Definition \ref{definition3.1}.

Moreovere, we in fact have defined a $(1,1)$ correspondence between the set $\{(M,k)\}$ of all polynomial-time deterministic Turing machines and $\mathbb{N}_1$ if any integer that cannot be decoded is deemed to represent the trivial polynomial-time deterministic Turing machine. This is analogous to the situation on p. 241 of \cite{Tur37}; hence the set $\{(M,k)\}$ of all polynomial-time deterministic Turing machines is enumerable.
\vskip 0.3 cm
\begin{remark}
\label{remark3.2}
There is another way to {\em enumerate} all polynomial-time deterministic Turing machines without encoding the order of polynomial-time deterministic Turing machines into their representation. It relies on the {\em Cantor pairing function} : $$\pi:\mathbb{N}\times\mathbb{N}\rightarrow\mathbb{N}$$ defined by $$\pi(k_1,k_2):=\frac{1}{2}(k_1+k_2)(k_1+k_2+1)+k_2$$ where $k_1,k_2\in\mathbb{N}$. Since the Cantor pairing function (see Figure \ref{5} below, which is from \cite{ANMOUS5}) is invertible (see \cite{ANMOUS5}), it is a bijection between $\mathbb{N}\times\mathbb{N}$ and $\mathbb{N}$. As we have shown that any polynomial-time deterministic Turing machine itself is an integer, we can place any polynomial-time deterministic Turing machine $M$ and its order $k$ in the tuple $(M,k)$ and use the Cantor pairing function to map the tuple $(M,k)$ to an integer in $\mathbb{N}_1$. Recall that there is a bijection between $\mathbb{N}$ and $\mathbb{N}_1$. Obviously, the inverse of such a Cantor pairing function is an enumeration of the set $\{(M,k)\}$. For more complete details, see \cite{Lin21a}.
\end{remark}

\begin{figure}[ht]
\centering
\includegraphics[width=8.3cm]{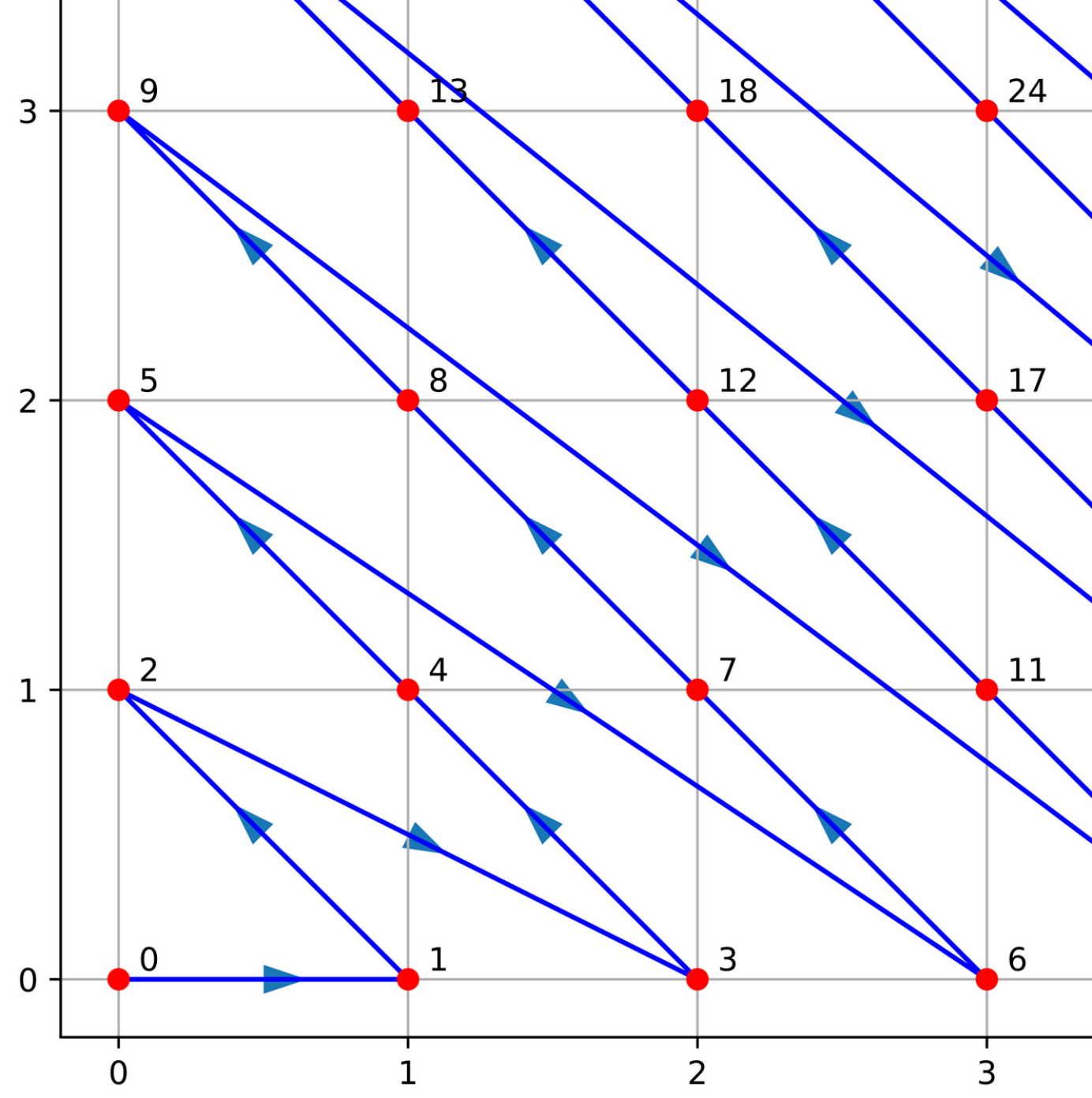}
\caption{\label{5}Cantor pairing function}
\end{figure}

\vskip 0.3 cm
\section{Diagonalization against Polynomial-Time Deterministic Turing Machines}
\label{sec:diagonalization}
\vskip 0.3 cm

The {\em diagonalization} technique is a powerful method for establishing space and time hierarchies on uniform computing models (i.e., the Turing machine model); see e.g., \cite{AB09, HS65, SHL65, AHU74, SFM78}. For more details on this technique, we refer the reader to Turing's original article \cite{Tur37} or the survey article \cite{For00}.

Our task is to design a probabilistic Turing machine $M_0$ that treats its input string $x$ both as an encoding of a tuple $(M,k)$ of a deterministic $n^k+k$ time-bounded Turing machine $M$ and as the input to $M$. One capability of $M_0$ is the ability to simulate any deterministic Turing machine given its specification. Before proceeding, suppose that $M_0$ has two specially designated states: the pretended accept state $q_{pa}$ and the pretended reject state $q_{pr}$, which play an important role in the following argument. We shall have $M_0$ determine whether the deterministic $T(n)=n^k+k$ time-bounded Turing machine $(M,k)$ accepts the input $x$ without using more than $O(T(n)\log T(n))$ steps (by Lemma \ref{lemma2.3}). If $M$ accepts $x$ within time $T(n)$, then $M_0$ moves to state $q_{pa}$. Otherwise, $M_0$ moves to state $q_{pr}$. Note that the states $q_{acc}$ and $q_{rej}$ are the actual accept state and the actual reject state of $M_0$,respectively; they will be added to $M_0$ at the end.

Let $\lambda\in (0,1)$ be a rational and $\epsilon>0$ be any rational. The final step of $M_0$ is as follows: if $M_0$ is in state $q_{pr}$, it transitions to $q_{acc}$ with probability $\lambda+\epsilon$ and to $q_{rej}$ with probability $1-(\lambda+\epsilon)$, then halts. Otherwise (i.e., if it is in $q_{pa}$), it transitions to $q_{acc}$ with probability $1-(\lambda+\epsilon)$ and to $q_{rej}$ with probability $\lambda+\epsilon$, then halts. This is illustrated in Figure \ref{fig1} below. 
    
\begin{figure}[ht]
\centering
\includegraphics[width=10cm]{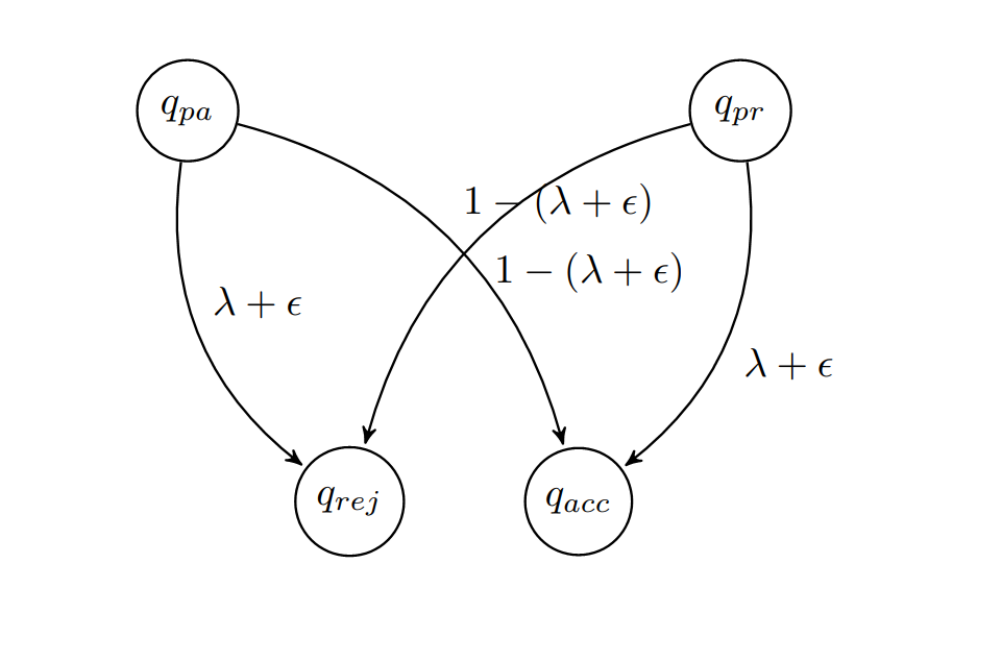}
\caption{\label{fig1}The probabilistic transitions of $M_0$ and the corresponding probabilities}
\end{figure}

Concretely, we will prove the following:

\begin{theorem}
\label{theorem4.1}
There exists a language $L_d$ that is accepted by a probabilistic Turing machine $M_0$ but by no polynomial-time deterministic Turing machine (i.e., $L_d\not\in \mathcal{P}$).
\end{theorem}

\begin{proof}
Let $M_0$ be a four-tape probabilistic Turing machine that operates as follows on an input string $x$ of length $n$.
\begin{enumerate}
       \item{$M_0$ decodes the tuple encoded by $x$. If $x$ is not the encoding of some single-tape polynomial-time deterministic Turing machine $\widehat{M}_j$ for some $j$, then GOTO (5); else determine $t$, the number of tape symbols used by $\widehat{M}_j$; $s$, its number of states; and $k$, its order of polynomial. The third tape of $M_0$ can be used as ``scratch" memory to calculate $t$.}
       \item{Then $M_0$ lays off on its second tape $n$ blocks of \[\lceil\log t\rceil\] cells each, the blocks being separated by a single cell holding a marker $\#$, i.e., there are $$(1+\lceil\log t\rceil)n$$ cells in all. Each tape symbol occurring in a cell of $\widehat{M}_j$'s tape will be encoded as a binary number in the corresponding block of the second tape of $M_0$. Initially, $M_0$ places $\widehat{M}_j$'s input, in binary coded form, in the blocks of tape $2$, filling the unused blocks with the code for the blank.}
       \item{On tape $3$, $M_0$ sets up a block of $$\lceil\log(n^{k+1}-2)\rceil$$ cells, initialized to all $0$'s. Tape $3$ is used as a counter to count up to $$n^{k+1}-2.\footnote{Suppose that $\widehat{M}_j$ is a deterministic $T(n)$ time-bounded Turing machine where $$T(n)=n^k+k.$$ By Lemma \ref{lemma2.3}, the simulation can be performed within time $T(n)\log T(n)$, which is less than $n^{k+1}-2$. There are $2$ additional steps (described by the transition rules (\ref{eq1}) and (\ref{eq2}) below) after the simulation completes, so we set the counter to count up to $n^{k+1}-2$ to ensure that $M_0$ runs within at most $n^{k+1}$ steps overall.}
       $$
       }
       \item{ $M_0$ simulates $\widehat{M}_j$ using tape $1$ (its input tape) to determine the moves of $\widehat{M}_j$ and tape $2$ to simulate the tape of $\widehat{M}_j$. The moves of $\widehat{M}_j$ are counted in binary in the block on tape $3$, and tape $4$ is used to hold the state of $\widehat{M}_j$. If $\widehat{M}_j$ accepts, then $M_0$ moves to state $q_{pa}$; if $\widehat{M}_j$ rejects, then $M_0$ moves to state $q_{pr}$. Moreover, if the counter on tape $3$ overflows (we must reject such inputs with probability $1$), $M_0$ moves to state $q_{rej}$ (note that the state $q_{rej}$ will be added in what follows).}
       \item {
       Since $x$ is not an encoding of a single-tape polynomial-time deterministic Turing machine, $M_0$ transitions to the actual reject state $q_{rej}$ with probability $1$ (the state $q_{rej}$ will be added into $M_0$ in what follows).
       }
\end{enumerate}

\vskip 0.3 cm
\begin{remark}
\label{remark4.1}
So far, the above design of $M_0$, is essentially a universal deterministic Turing machine (and hence also a probability one, since every deterministic machine is a special case of a probabilistic machine). It simply simulates any polynomial-time deterministic Turing machine and performs no diagonalization operation.
\end{remark}
\vskip 0.3 cm

To complete the construction, we add some additional probabilistic transition rules to $M_0$: For any $4$-tuple $X=(a_1,a_2,a_3,a_4)\in \{0,1\}\times\{0,1,\#\}\times\{0,1\}^2$,\footnote{Note that the input $w$ to $M_0$ is a string in $\{0,1\}^*$ that encodes a single-tape polynomial-time deterministic Turing machine, but the tape alphabet of the second tape is $\{0,1,\#\}$.} where none of the $a_i$ is the blank symbol $\mathrm{b}$:
\begin{equation}
\label{eq1}
\begin{split}
          \delta(q_{pa},X)=&\,\, (q_{pa},(a_1,S),(\mathrm{b},S),(\mathrm{b},S),(a_4,S))\quad\text{ with probability $1$ }\\
          \delta(q_{pr},X)=&\,\,(q_{pr},(a_1,S),(\mathrm{b},S),(\mathrm{b},S),(a_4,S))\quad\text{ with probability $1$. }\\
\end{split}
\end{equation}

Note that $M_0$'s tape alphabet is $\{0,1,\mathrm{b},\#\}$, where $\#$ appears on tape $2$. These probabilistic transition rules (given by (\ref{eq1})) ensure that when $M_0$ is in $q_{pa}$ or $q_{pr}$, it keeps the current state unchanged, replaces the tuple of tape symbols $$X=(a_1,a_2,a_3,a_4)\in\{0,1\}\times\{0,1,\#\}\times\{0,1\}^2$$ with $(a_1,\mathrm{b},\mathrm{b},a_4)$, and leaves the heads stationary. We assume that upon entering $q_{pa}$ or $q_{pr}$, the heads are scanning symbols $0$ or $1$ (not the blank $\mathrm{b}$); this is easy to achieve.

\vskip 0.3 cm
\begin{remark}
\label{remark4.2}
In the rules above, $M_0$ does not rewrite the contents of tape $1$ or tape $4$. Tape 1 is read-only input tape (constant for a fixed input), and tape $4$ holds the state of $\widehat{M}_j$ (also constant during the simulation of a fixed input). We choose not to overwrite them with blanks to make $M_0$ appear more well-behaved.
\end{remark}
\vskip 0.3 cm

Let $\lambda=\frac{2}{3}$ and $\epsilon>0$ be an arbitrary small rational. We now add the two actual halting states $q_{acc}$ and $q_{rej}$ to $M_0$ (when $M_0$ enters one of these states, it halts with no further transitions). Since the simulation of any fixed input has already been completed, we let $M_0$ perform the diagonalization probabilistically by adding the following probabilistic transition rules. For any $X=(a_1,\mathrm{b},\mathrm{b},a_4)$ where $a_1,a_4\in\{0,1\}$:
\begin{equation}
\label{eq2}
\begin{split}
          \delta(q_{pa},a_1,\mathrm{b},\mathrm{b},a_4)=& (q_{rej},(a_1,S),(\mathrm{b},S),(\mathrm{b},S),(a_4,S))\\
          &\quad\text{ with probability $\frac{2}{3}+\epsilon$}\\
          \delta(q_{pa},a_1,\mathrm{b},\mathrm{b},a_4)=& (q_{acc},(a_1,S),(\mathrm{b},S),(\mathrm{b},S),(a_4,S))\\
          &\quad\text{ with probability $\frac{1}{3}-\epsilon$}\\
          \delta(q_{pr},a_1,\mathrm{b},\mathrm{b},a_4)=&(q_{acc},(a_1,S),(\mathrm{b},S),(\mathrm{b},S),(a_4,S))\\
          &\quad\text{ with probability $\frac{2}{3}+\epsilon$}\\
          \delta(q_{pr},a_1,\mathrm{b},\mathrm{b},a_4)=&(q_{rej},(a_1,S),(\mathrm{b},S),(\mathrm{b},S),(a_4,S))\\
          &\quad\text{ with probability $\frac{1}{3}-\epsilon$.}
\end{split}
\end{equation}

\vskip 0.3 cm
\begin{remark}
\label{remark4.3}
After these additions, $M_0$ is no longer a universal deterministic Turing machine, but it remains a probabilistic Turing machine that can simulate any polynomial-time deterministic Turing machines and flip its answers probabilistically (i.e., perform diagonalization). Most importantly, $M_0$ does not appear in the enumeration $e$, because its probabilistic transitions (with probabilities strictly between $0$ and $1$) are not encoded in the same way as the polynomial-time deterministic Turing machines described in Section \ref{sec:enumeration}.
\end{remark}
\vskip 0.3 cm

The probabilistic Turing machine $M_0$ described above is of time complexity $S(n)$ (currently unknown). By Lemma \ref{lemma2.1} (regarding a probabilistic Turing machine as a nondeterministic one with a probability distribution over the next-move relation), $M_0$ is equivalent to a single-tape probabilistic $O(S^2(n))$ time-bounded Turing machine, and it of course accepts some language $L_d$.

For any $x\in L_d$, $M_0$ accepts $x$ with probability at least $\frac{2}{3}+\epsilon\geq\frac{2}{3}$. For any $x\not\in L_d$, $M_0$ accepts $x$ with probability at most $\frac{1}{3}-\epsilon\leq\frac{1}{3}$. See Figure \ref{2} below for clarity on the acceptance conditions.
     
\begin{figure}[ht]
\centering
\includegraphics[width=10cm]{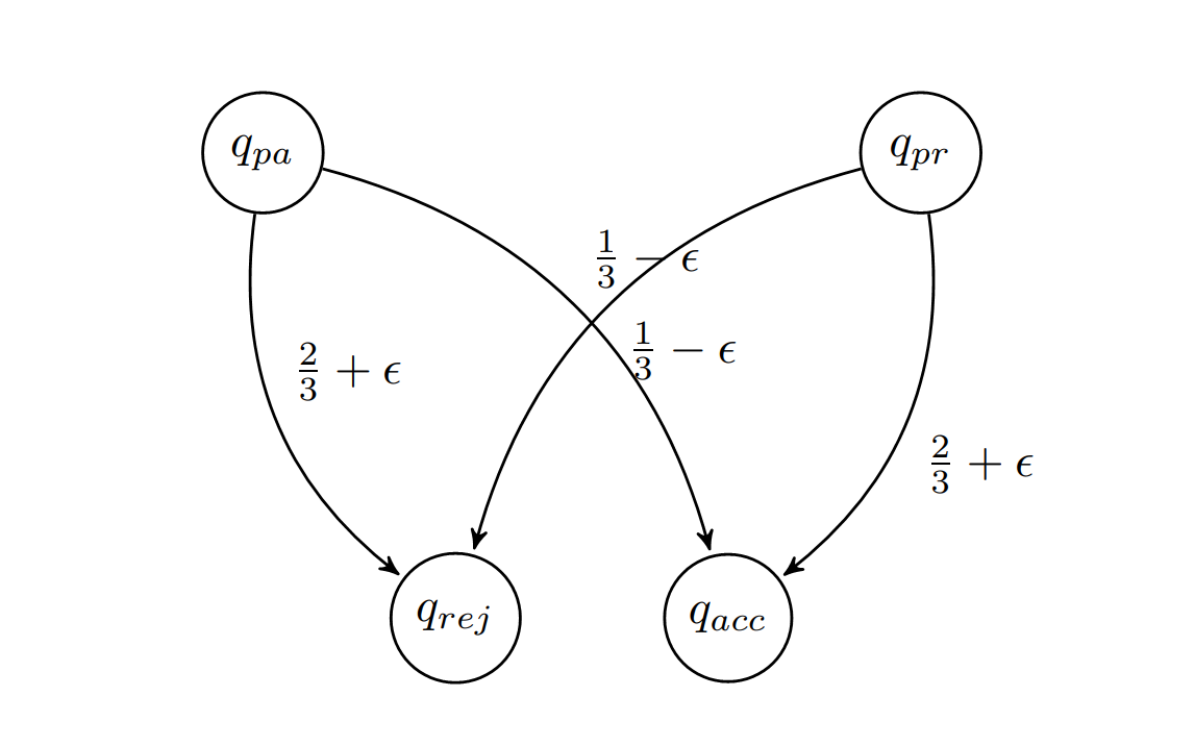}
\caption{\label{2}The transitions of states of $M_0$ and corresponding probabilities by setting $\lambda=\frac{2}{3}$}
\end{figure}

Suppose, for the sake of contradiction, that $L_d$ is accepted by some deterministic Turing machine in the enumeration $e$, say the $i$-th deterministic Turing machine, which is a deterministic $T(n)=n^k+k$ time-bounded Turing machine $\widehat{M}_i$. By Lemma \ref{lemma2.1} we may assume that $\widehat{M}_i$ is a single-tape deterministic Turing machine. Let $\widehat{M}_i$ have $s$ states and $t$ tape symbols. Since $\widehat{M_i}$ appears infinitely often in the enumeration,\footnote{We know that we may prefix $1$s at will to find larger and larger integers representing the same set of quintuples of the same deterministic Turing machine $M_i$; thus, there are infinitely many binary strings of sufficient length that represent deterministic Turing machine $M_i$.} we have
$$\aligned
\lim_{n\rightarrow\infty}&\frac{T(n)\log T(n)}{n^{k+1}-2}\\
         =&\lim_{n\rightarrow\infty}\frac{(n^k+k)\log(n^k+k)}{n^{k+1}-2}\\
         =&\lim_{n\rightarrow\infty}\Big(\frac{n^k\log(n^k+k)}{n^{k+1}-2}+\frac{k\log(n^k+k)}{n^{k+1}-2}\Big)\\
         =&0\\
         <&1.
\endaligned$$

So, there exists an $N_0>0$ such that for any $N\geq N_0$, $$T(N)\log T(N)<N^{k+1}-2,$$ which implies that for a sufficiently long $w$, say $|w|\geq N_0$, and $M_w$ denoted by such $w$ is $\widehat{M}_i$, we have that $$T(|w|)\log T(|w|)<|w|^{k+1}-2.$$

Thus, on input $w$, $M_0$ has sufficient time to simulate $M_w$ and reaches the state $q_{pr}$ or $q_{pa}$. After running the additional $2$ steps (i.e., the probabilistic transition rules (\ref{eq1}) and (\ref{eq2})) to finish the diagonalization operation probabilistically, $M_0$ accepts with probability $\geq\frac{2}{3}$ if and only if $M_w$ rejects and accepts with probability $\leq\frac{1}{3}$ if and only if $M_w$ accepts, which further means that $w\in L_d$ if and only if $M_w$ rejects $w$ and $w\not\in L_d$ if and only if $M_w$ accepts $w$. But we assumed that $\widehat{M}_i$ accepted $L_d$, i.e., $\widehat{M}_i$ agreed with $M_0$ on all inputs. This is a contradiction. We thus conclude that $\widehat{M}_i$ does not exist; we obtain $$L_d\not\in \mathcal{P}. $$
\end{proof}

\vskip 0.3 cm

Next, we are going to show that the probabilistic Turing machine $M_0$ runs in time $O(n^k)$ for all $k\in\mathbb{N}_1$:

\begin{theorem}
\label{theorem4.2}
The probabilistic Turing machine $M_0$ constructed in proof of Theorem \ref{theorem4.1} runs in time $O(n^k)$ for any $k\in\mathbb{N}_1$.
\end{theorem}

\begin{proof}
The quickest way to show this theorem is to prove that for any input $w$ to $M_0$, there is a corresponding positive integer $i_w\in\mathbb{N}_1$ such that $M_0$ runs at most $|w|^{i_w+1}$ steps.

On the one hand, if the input $x$ encodes a deterministic $T(n)$ time-bounded Turing machine, say $T(n)=O(n^k)$, then $M_0$ runs at most $$|x|^{k+1}$$ steps by the construction (the simulation can be completed in $|x|^{k+1}-2$, and to halt itself, there are $2$ additional steps to run). This means that $k$ is the required positive integer (i.e., $i_w=k$) in this case. This holds true for all polynomial-time deterministic Turing machines as input to $M_0$ with $k$ the order of the corresponding polynomial-time deterministic Turing machine.

But on the other hand, if the input $x$ does not encode some polynomial-time deterministic Turing machine, then it rejects, and the running time of $M_0$ is within $O(|x|)$ (i.e., $M_0$ runs at most $|x|^2$ steps), which means $1$ is the required positive integer. So $M_0$ is a probabilistic $$S(n)=\max\{n^{k+1},n^2\}$$ time-bounded Turing machine for any $k\in\mathbb{N}_1$. That is, $M_0$ is a probabilistic \[O(n^k)\] time-bounded Turing machine for all $k\in\mathbb{N}_1$.
\end{proof}

\vskip 0.3 cm
\section{$L_d$ is in $\mathcal{BPP}$}
\label{sec:l_dinbpp}
\vskip 0.3 cm

Next, we show that the language $L_d$ is in the class $\mathcal{BPP}$. 

The proof of $L_d\in\mathcal{BPP}$ is essentially the same as the proof that the diagonalization language in \cite{Lin21a} belongs to $\mathcal{NP}$ (see Section $5$ in \cite{Lin21a}). For completeness and simplicity, we present the simpler proof below:
\begin{theorem}
\label{theorem5.1}
The language $L_d$ accepted by the probabilistic Turing machine $M_0$ is in $\mathcal{BPP}$.
\end{theorem}
\begin{proof}
We first define the family of languages $$\{L_d^i\}_{i\in\mathbb{N}_1}$$ by adding a new tape to $M_0$ as a counter to count up to $$n^{i+1},$$ which means that $M_0$ turns itself off when $M_0$ continues to run $2$ steps after the counter of tape $3$ exceeds $n^{k+1}-2$ or when the counter of the newly added tape exceeds $n^{i+1}$:
$$\aligned
L_d^i\overset{\rm def}{=}&\mbox{ language accepted by $M_0$ running within time $O(n^i)$ for fixed $i\in\mathbb{N}_1$.}\\
         &\mbox{ That is, $M_0$ turns itself off mandatorily when its moves made by $M_0$ during }\\
         &\mbox{ the computation exceed $n^{i+1}$ steps.}
\endaligned$$

 Then by construction, $M_0$ runs at most $|w|^{i_w+1}$ steps for any input $w$ where $i_w\in\mathbb{N}_1$ (i.e., $M_0$ runs within time $O(n^i)$ for any $i\in\mathbb{N}_1$, see Theorem \ref{theorem4.2}); we thus have
\begin{equation}
\label{eq3}
\begin{split}
L_d=&\bigcup_{i\in\mathbb{N}_1}L_d^i.
\end{split}
\end{equation}

Furthermore, $$L_d^i\subseteq L_d^{i+1},\quad\mbox{for each fixed $i\in\mathbb{N}_1$} $$ since for any word $w\in L_d^i$ accepted by $M_0$ within $O(n^i)$ steps, it surely can be accepted by $M_0$ within $O(n^{i+1})$ steps, i.e., $$w\in L_d^{i+1}. $$

This gives that for any fixed $i\in\mathbb{N}_1$,
\begin{equation}
\label{eq4}
\begin{split}
L_d^1\subseteq L_d^2\subseteq\cdots\subseteq L_d^i\subseteq L_d^{i+1}\subseteq\cdots
\end{split}
\end{equation}

Now, we assume that $$L_d\not\in\mathcal{BPP},$$ then there must exist at least a fixed $i\in\mathbb{N}_1$ such that $$L_d^i\not\in\mathcal{BPP}. $$ But by definition (Definition \ref{definition2.9}), $L_d^i$ is the language accepted by the probabilistic Turing machine $M_0$ running within time $n^{i+1}+(i+1)$, i.e., $$L_d^i\in{\rm BPTIME}[n^i],$$ which is clearly a contradiction. We thus can claim that such an $i$ can not be found. Equivalently,
\begin{equation}
\label{eq5}
\begin{split}
L_d^i\in\mathcal{BPP}\quad\text{ for all $i\in\mathbb{N}_1$,}
\end{split}
\end{equation}
which further implies
$$
L_d\in\mathcal{BPP},
$$
as required.
\end{proof}

\vskip 0.3 cm
Now we are ready to present the proof of Theorem \ref{theorem1.1}:

\vskip 0.2 cm
\indent{\it Proof of Theorem \ref{theorem1.1}}. It is immediate that Theorem \ref{theorem1.1} follows from Theorem \ref{theorem4.1} and Theorem \ref{theorem5.1}. \Q.E.D

\section{Proof of $\mathcal{P}\subsetneqq\mathcal{RP}$}
\label{sec:proof_of_pneqrp}
\vskip 0.3 cm

The proof of Theorem \ref{theorem1.3} is also divided into two auxiliary theorems:

\begin{theorem}
\label{theorem6.1}
There exists a language $\widetilde{L_d}$ that is accepted by a probabilistic Turing machine $M'_0$ but by no polynomial-time deterministic Turing machine (i.e., $\widetilde{L_d}\not\in \mathcal{P}$).
\end{theorem}

\begin{theorem}
\label{theorem6.2}
The language $\widetilde{L_d}$ accepted by the probabilistic Turing machine $M'_0$ is in $\mathcal{RP}$.
\end{theorem}

To prove Theorem \ref{theorem6.1}, we can reuse the construction of the probabilistic Turing machine $M_0$ from the proof of Theorem \ref{theorem4.1}, except for the final step of $M_0$. We assume that $M'_0$ is identical to $M_0$ except for its final step. In this final step, the probabilistic transition rules of $M_0$ given by (\ref{eq2}) are replaced with the following rules:

Let $\epsilon>0$ be any small positive real number.

\begin{equation}
\label{eq6}
\begin{split}
\delta(q_{pa},a_1,\mathrm{b},\mathrm{b},a_4)=& (q_{rej},(a_1,S),(\mathrm{b},S),(\mathrm{b},S),(a_4,S))\\
          &\quad\text{ with probability $1$}\\
          \delta(q_{pa},a_1,\mathrm{b},\mathrm{b},a_4)=& (q_{acc},(a_1,S),(\mathrm{b},S),(\mathrm{b},S),(a_4,S))\\
          &\quad\text{ with probability $\epsilon$}\\
          \delta(q_{pr},a_1,\mathrm{b},\mathrm{b},a_4)=&(q_{acc},(a_1,S),(\mathrm{b},S),(\mathrm{b},S),(a_4,S))\\
          &\quad\text{ with probability $\frac{1}{2}+\epsilon$}\\
          \delta(q_{pr},a_1,\mathrm{b},\mathrm{b},a_4)=&(q_{rej},(a_1,S),(\mathrm{b},S),(\mathrm{b},S),(a_4,S))\\
          &\quad\text{ with probability $\frac{1}{2}-\epsilon$}
\end{split}
\end{equation}
for any $X=(a_1,\mathrm{b},\mathrm{b},a_4)$ where $a_1,a_4\in\{0,1\}$.

That is, if $M'_0$ is in state $q_{pr}$, it transitions to $q_{acc}$ with probability $\frac{1}{2}+\epsilon$ and to $q_{rej}$ with probability $1-(\frac{1}{2}+\epsilon)$, then halts. Otherwise (i.e., if it is in $q_{pa}$), it transitions to $q_{rej}$ with probability $1-\epsilon$ and to $q_{acc}$ with probability $\epsilon$, then halts. This is illustrated in Figure \ref{3} below. We denote by $\widetilde{L_d}$ the language accepted by $M'_0$ in the limit as $\epsilon$ tends to $0$.  

\begin{figure}[ht]
\centering
\includegraphics[width=10cm]{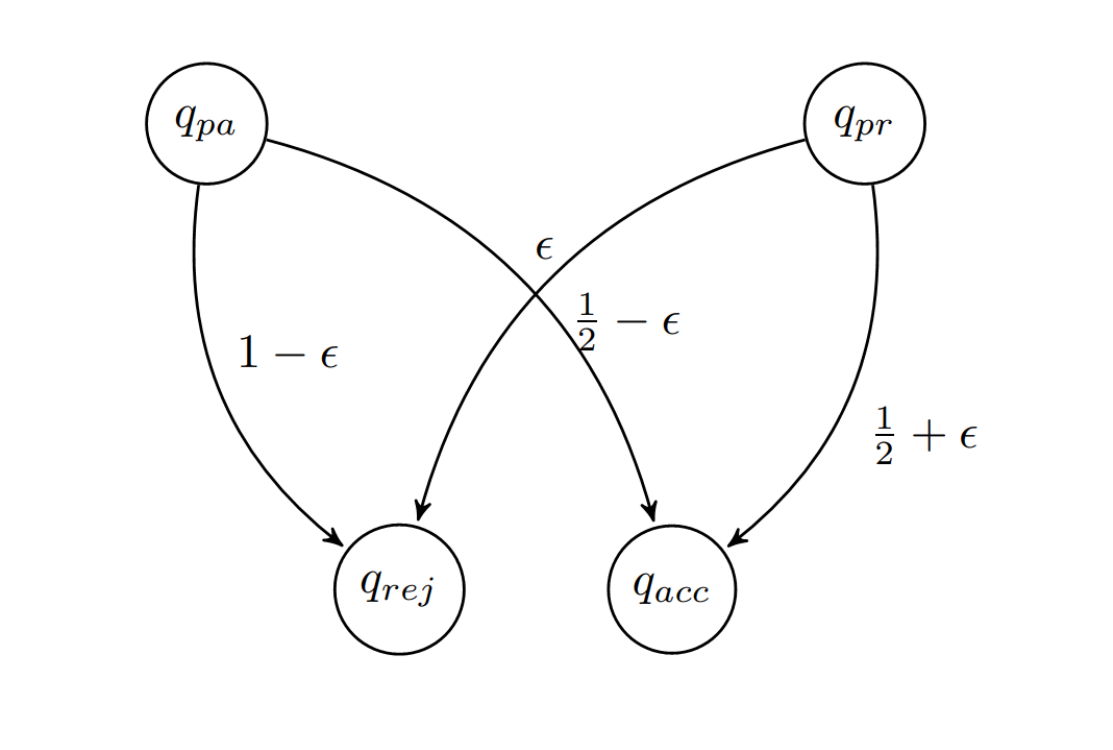}
\caption{\label{3}The transitions of states and corresponding probabilities for $\widetilde{L_d}\in\mathcal{RP}$}
\end{figure}

Now, we  see that for $x\in\widetilde{L_d}$, $$\mathrm{Pr}[M'_0(x)\,\,{\rm accepts}]\geq\frac{1}{2}, $$ and for $x\not\in\widetilde{L_d}$, $$\mathrm{Pr}[M'_0(x)\,\,{\rm accepts}]=0.$$ Moreover, by arguments similar to those in the proof of Theorem \ref{theorem4.1}, it is clear that $$ \widetilde{L_d}\not\in\mathcal{P}. $$ This completes the proof of Theorem \ref{theorem6.1}. \Q.E.D

The remainder of the proof that $\widetilde{L_d}\in\mathcal{RP}$ is similar to the proof of Theorem \ref{theorem5.1}.\\
\indent{\em Proof of Theorem \ref{theorem6.2}:} We first define the family of languages $$\{\widetilde{L_d^i}\}_{i\in\mathbb{N}_1}$$ by adding a new tape to $M'_0$ as a counter to count up to $$n^{i+1}.$$ This means that $M'_0$ turns itself off when it continues to run $2$ more steps after the counter of tape $3$ exceeds $n^{k+1}-2$, or when the counter of the newly added tape exceeds $n^{i+1}$:
$$\aligned
\widetilde{L_d}^i\overset{\rm def}{=}&\mbox{ language accepted by $M'_0$ running within time $O(n^i)$ for fixed $i\in\mathbb{N}_1$.}\\
         &\mbox{ That is, $M'_0$ turns itself off mandatorily when its moves made by $M'_0$ during }\\
         &\mbox{ the computation exceed $n^{i+1}$ steps.}
\endaligned$$

Next, similar to the proof of Theorem \ref{theorem5.1}, we can show that
$$
\widetilde{L_d}=\bigcup_{i\in\mathbb{N}_1}\widetilde{L_d^i},
$$ 
$$
\widetilde{L_d^1}\subseteq \widetilde{L_d^2}\subseteq\cdots\subseteq \widetilde{L_d^i}\subseteq \widetilde{L_d^{i+1}}\subseteq\cdots,
$$and
$$
\widetilde{L_d^i}\in{\rm RPTIME}[n^i]\quad\text{ for all $i\in\mathbb{N}_1$.}
$$ From the above arguments it immediate follows that $$\widetilde{L_d}\in\mathcal{RP}.$$\Q.E.D

\vskip 0.3 cm
\section{Proof of $\mathcal{P}\subsetneqq{\rm co}\mathcal{RP}$}
 \label{sec:proof_of_pneqcorp}
\vskip 0.3 cm

Similarly, the proof of Theorem \ref{theorem1.5} is divided into two auxiliary theorems:

\begin{theorem}
\label{theorem7.1}
There exists a language $\widehat{L_d}$ that is accepted by a probabilistic Turing machine $N'_0$ but by no polynomial-time deterministic Turing machine (i.e., $\widehat{L_d}\not\in \mathcal{P}$).
\end{theorem}

\begin{theorem}
\label{theorem7.2}
The language $\widehat{L_d}$ accepted by the probabilistic Turing machine $N'_0$ is in ${\rm co}\mathcal{RP}$.
\end{theorem}

To prove Theorem \ref{theorem7.1}, we reuse the construction of the probabilistic Turing machine $M_0$ from the proof of Theorem \ref{theorem4.1}, except for its final step. We assume that $N'_0$ is identical to $M_0$ except for its final step. In addition, we introduce a new state $q_{uk}$ ( ``unknown") in $N'_0$. When $N'_0$ enters state $q_{uk}$, it print a symbol representing ``unknown" on the output tape for the given input and then halts. In this final step, the probabilistic transition rules of $M_0$ given by (\ref{eq2}) are replaced with the following rules:

Let $\epsilon>0$ be any small positive real number.

\begin{equation}
\label{eq7}
\begin{split}
 \delta(q_{pa},a_1,\mathrm{b},\mathrm{b},a_4)=& (q_{rej},(a_1,S),(\mathrm{b},S),(\mathrm{b},S),(a_4,S))\\
 &\quad\text{ with probability $\frac{1}{2}-\epsilon$}\\
 \delta(q_{pa},a_1,\mathrm{b},\mathrm{b},a_4)=& (q_{uk},(a_1,S),(\mathrm{b},S),(\mathrm{b},S),(a_4,S))\\
 &\quad\text{ with probability $\frac{1}{2}$}\\
 \delta(q_{pa},a_1,\mathrm{b},\mathrm{b},a_4)=& (q_{acc},(a_1,S),(\mathrm{b},S),(\mathrm{b},S),(a_4,S))\\
 &\quad\text{ with probability $\epsilon$}\\
 \delta(q_{pr},a_1,\mathrm{b},\mathrm{b},a_4)=&(q_{acc},(a_1,S),(\mathrm{b},S),(\mathrm{b},S),(a_4,S))\\
 &\quad\text{ with probability $1-\epsilon$}\\
 \delta(q_{pr},a_1,\mathrm{b},\mathrm{b},a_4)=&(q_{rej},(a_1,S),(\mathrm{b},S),(\mathrm{b},S),(a_4,S))\\
  &\quad\text{ with probability $\epsilon$}
  \end{split}
 \end{equation}
 for any $X=(a_1,\mathrm{b},\mathrm{b},a_4)$ where $a_1,a_4\in\{0,1\}$.

That is, if $N'_0$ is in state $q_{pa}$, it transitions to $q_{acc}$ with probability $\epsilon$, to $q_{rej}$ with probability $\frac{1}{2}-\epsilon$, and to $q_{uk}$ with probability $\frac{1}{2}$, and then halts. Otherwise (i.e., if it is in $q_{pr}$), it transitions to $q_{acc}$ with probability $1-\epsilon$ and to $q_{rej}$ with probability $\epsilon$, and then halts. This is illustrated in Figure \ref{4} below. We denote by $\widehat{L_d}$ the language accepted by $N'_0$ in the limit as $\epsilon$ tends to $0$.  

\begin{figure}[ht]
\centering
\includegraphics[width=10.5cm]{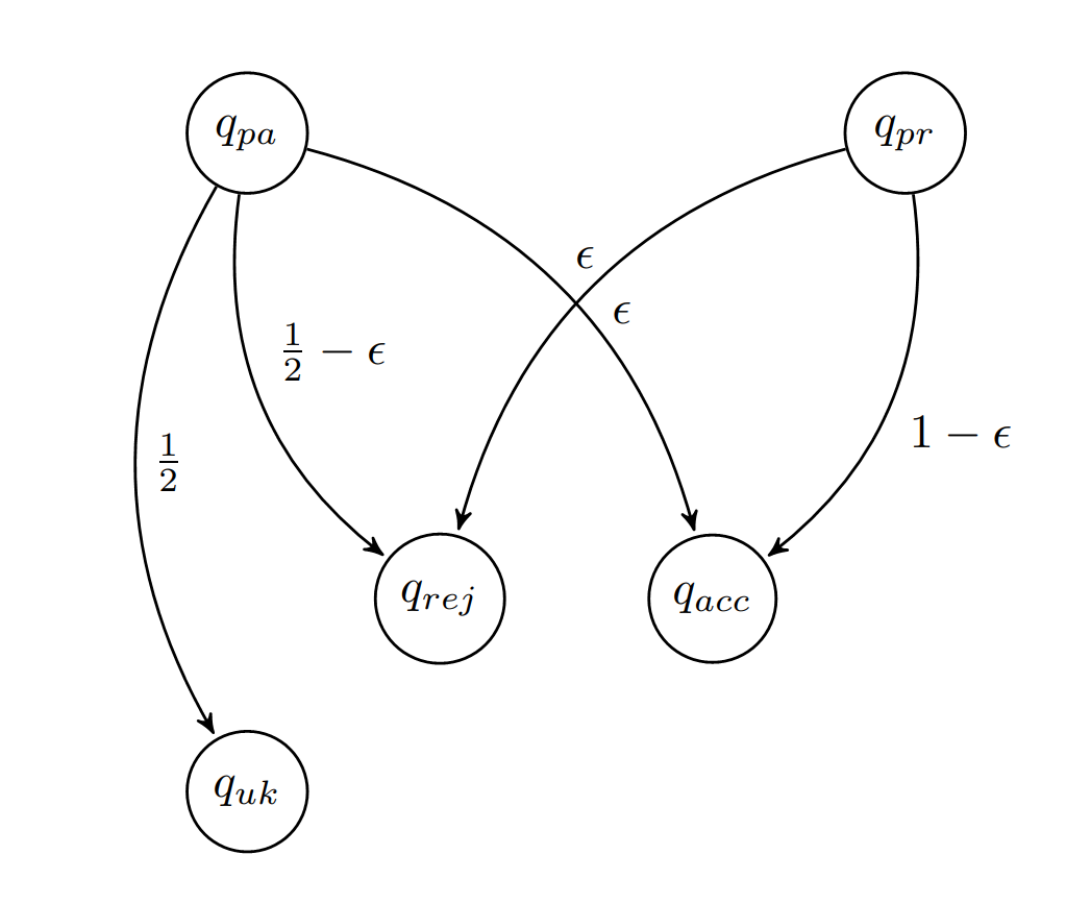}
\caption{\label{4}The transitions of states of $N'_0$ and corresponding probabilities for $\widehat{L_d}\in{\rm co}\mathcal{RP}$}
\end{figure}

Now, for $x\in\widehat{L_d}$, $$\mathrm{Pr}[N'_0(x)\,\,{\rm accepts}]= 1,$$ and for $x\not\in\widehat{L_d}$,
$$
  \mathrm{Pr}[N'_0(x)\text{ accepts}]\leq\frac{1}{2}.
$$ Moreover, by arguments similar to those in the proof of Theorem \ref{theorem4.1}, it is clear that $$ \widehat{L_d}\not\in\mathcal{P}. $$ This completes the proof of Theorem \ref{theorem7.1}. \Q.E.D

The remainder of the proof that $\widehat{L_d}\in{\rm co}\mathcal{RP}$ is similar to the proof of Theorem \ref{theorem5.1}.\\

\indent{\em Proof of Theorem \ref{theorem7.2}:} We first define the family of languages $$\{\widehat{L_d^i}\}_{i\in\mathbb{N}_1}$$ by adding a new tape to $N'_0$ as a counter that counts up to $$n^{i+1}.$$ This means that $N'_0$ turns itself off when it continues to run $2$ more steps after the counter on tape $3$ exceeds $n^{k+1}-2$, or when the counter on the newly added tape exceeds $n^{i+1}$:
$$\aligned
\widehat{L_d}^i\overset{\rm def}{=}&\mbox{ language accepted by $N'_0$ running within time $O(n^i)$ for fixed $i\in\mathbb{N}_1$.}\\
         &\mbox{ That is, $N'_0$ turns itself off mandatorily when its moves made by $N'_0$ during }\\
         &\mbox{ the computation exceed $n^{i+1}$ steps.}
\endaligned$$

Next, similar to the proof of Theorem \ref{theorem5.1}, we can show that
$$
\widehat{L_d}=\bigcup_{i\in\mathbb{N}_1}\widehat{L_d^i},
$$ 
$$
\widehat{L_d^1}\subseteq \widehat{L_d^2}\subseteq\cdots\subseteq \widehat{L_d^i}\subseteq \widehat{L_d^{i+1}}\subseteq\cdots,
$$and
$$
\widehat{L_d^i}\in{\rm coRPTIME}[n^i]\quad\text{ for all $i\in\mathbb{N}_1$.}
$$ From the above arguments it immediate follows that $$\widehat{L_d}\in{\rm co}\mathcal{RP}.$$\Q.E.D
 
\vskip 0.5 cm
\begin{remark}
\label{remark7.1}
 As a matter of fact, $\widehat{L_d}$ is also in $\mathcal{RP}$. For $x\in\widehat{L_d}$, $$\mathrm{Pr}\text{[}N'_0\text{ accepts }x\text{]}= 1\geq\frac{1}{2}, $$ and for $x\not\in\widehat{L_d}$, $$\mathrm{Pr}\text[N'_0\text{ accepts }x\text{]}= 0. $$ This also adheres to the definition of $\mathcal{RP}$. Thus, we can conclude that $$\widehat{L_d}\in\mathcal{RP}\cap{\rm co}\mathcal{RP},$$ which finishes the proof of Corollary \ref{corollary1.7}.
\end{remark}

\vskip 0.3 cm
\section{Randomness is Necessary (And Powerful)}
\label{sec:randomness_helps}
\vskip 0.3 cm

Since the $1970$s, scientists in the {\it theoretical computer science} community have been wondering how necessary the use of randomness is in algorithm applications. Today, randomness has become a very important tool in the design of efficient algorithms for many problems. Indeed, numerous problems for which no efficient deterministic algorithm were known have been solved efficiently by probabilistic algorithms, albeit with aa some small probability of error (which can be efficiently reduced). But is randomness essential, or can it be removed? And what quality of randomness is needed for the success of probabilistic algorithms?

Probabilistic algorithms are often the simplest or most efficient solutions to a given problem \cite{ACR98}.

In this section, we prove our main results showing that randomness cannot be eliminated --- i.e., that it plays an important and indispensable role in the design of probabilistic algorithms.

\subsection{Eliminating Randomness by Enumeration Method}
\vskip 0.3 cm

A common approach toward proving $$\mathcal{P}=\mathcal{BPP}$$ relies on derandomization, i.e., the design of general methods that allow efficient deterministic simulation of probabilistic algorithms. It is always possible to remove randomness at the cost of at most an exponential slowdown (see e.g., \cite{Vad12}), which implies that $$\mathcal{BPP}\subseteq\mathcal{EXP}.$$ As mentioned earlier, the proof of $$\mathcal{BPP}\subseteq\mathcal{EXP}$$ uses the enumeration method, but this method is infeasible in practice because it requires exponential time. However, if the algorithm uses only a small number of random bits, enumeration becomes feasible, as shown by the following proposition:

\begin{proposition}[cf. \cite{Vad12}, Proposition 3.3]
\label{proposition1}
 If $L$ has a probabilistic polynomial-time algorithm that runs in time $t(n)$ and uses $m(n)$ random bits, then $$L\in\textit{\em DTIME}[t(n)\times2^{m(n)}]. $$ In particular, if $t(n)$ is a polynomial and $$ m(n)=O(\log n), $$ then $$ L\in\mathcal{P}. $$
\end{proposition}

\begin{proof}
  See p. $51$--$52$ in \cite{Vad12}.
\end{proof}

\vskip 0.2 cm
As can be seen from Proposition \ref{proposition1}, one way to resolve the conjecture of $$\mathcal{P}=\mathcal{BPP}$$ is as follows: first show that the number of random bits used by any $\mathcal{BPP}$ algorithm can be reduced from polynomially many (poly$(n)$) to $O(\log n)$, and then eliminate randomness entirely by enumeration.

\vskip 0.2 cm
We are now ready to prove Theorem \ref{theorem1.8}.

\vskip 0.2 cm
\noindent{\em Proof of Theorem \ref{theorem1.8}.} We proceed by contradiction. Suppose that a probabilistic algorithm for the language $L_d$ uses only $O(\log n)$ random bits. Then, by Proposition \ref{proposition1}, we would have $$L_d\in\mathcal{P},$$ which contradicts Theorem \ref{theorem1.1}. This completes the proof. \Q.E.D

\vskip 0.3 cm
\subsection{Derandomize Probabilistic Algorithms by PRGs}
\vskip 0.3 cm

One of the two fundamental methods in the theory of derandomization is the use of {\em pseudorandom generators}, as shown by the following theorem, which exhibits how a {\em complexity-theoretic PRG} can be used to derandomize probabilistic algorithms.

\begin{definition}[\cite{NW94}]
\label{definition8.1}
  $G=\{G_n:\{0,1\}^{l(n)}\rightarrow\{0,1\}^n\}$, denoted by $G:l\rightarrow n$, is called a {\em pseudorandom generator} if for any circuit $C$ of size $n$:
  $$
  |\mathit{Pr}[C(y)]=1]-\mathit{Pr}[C(G(x))=1]|<1/n,
  $$
   where $y$ is chosen uniformly in $\{0,1\}^n$, and $x$ in $\{0,1\}^l$.
\end{definition}

\vskip 0.3 cm
\begin{theorem}[\cite{NW94}]
\label{theorem8.1}
  If there is a (complexity-theoretic) pseudorandom generator
  $$
  G: \{0,1\}^{l(t)}\rightarrow\{0,1\}^t,
  $$
  then
  $$
  {\rm BPTIME}[t(n)]\subseteq{\rm DTIME}[2^{O(l(t^2(n)))}].
  $$
\end{theorem}

\begin{proof}
  See \cite{NW94}.
\end{proof}

\vskip 0.2 cm
A direct corollary of the above theorem is the following:
\begin{corollary}
\label{corollary8.3}
  If there is a (complexity-theoretic) pseudorandom generator

  $$
  G:\{0,1\}^{l(t)}\rightarrow\{0,1\}^t
  $$
  with $l(t)=O(\log t)$, then
  $$
  \mathcal{P}=\mathcal{BPP}.
  $$
\end{corollary}
\begin{proof}
  Let $L\in\mathcal{BPP}$. Then $L\in\text{BPTIME}[n^k]$ for some fixed $k\in\mathbb{N}_1$. By Theorem \ref{theorem8.1}, we have
  $$\aligned
L\in&\,\,{\rm DTIME}[2^{O(l(n^{2k}))}]\\
  =&\,\,{\rm DTIME}[n^{O(1)}],\quad\text{(by $l(n^{2k})=O(2k\log n)$)}\\
  \subset&\,\,\mathcal{P}.
\endaligned
$$
\end{proof}

We are now ready to prove Theorem \ref{theorem1.10}:
\vskip 0.2 cm

\noindent{\em Proof of Theorem \ref{theorem1.10}.} Suppose, for the sake of contradiction, that there exists a (complexity-theoretic) pseudorandom generator:
$$
G:\{0,1\}^{l(t)}\rightarrow\{0,1\}^t
$$
with
$$
l(t)=O(\log t).
$$
Then, by Corollary \ref{corollary8.3}, we would have that
$$
L_d\in\mathcal{P},
$$
which contradicts Theorem \ref{theorem1.1}. This completes the proof.\Q.E.D

\vskip 0.3 cm
\subsection{Derandomize Probabilistic Algorithms by HSGs}
\vskip 0.3 cm

In addition to {\em pseudorandom generators}, {\em hitting set generators} (HSGs) constitute another important method in the theory of derandomization \cite{ACR98}.

\begin{definition}[\cite{ACR98}]
\label{definition8.2}
  A {\em hitting set generator} (HSG) is a function $H=\{H_n:\{0,1\}^{k(n)}\rightarrow \{0,1\}^n, n>0\}$ (denoted by $H: k(n)\rightarrow n$) that, for any sufficiently large $n$ and for any $n$-input boolean circuit $C$ with size at most $n$ such that
  $$
  \mathit{Pr}(C(\vec{y})=1)\geq\frac{1}{n},
  $$
  is required to provide {\em just} one ``example" $\vec{y}$ for which $C(\vec{y})=1$, that is, there exists $\vec{x}\in\{0,1\}^{k(n)}$ such that $C(H_n(\vec{x}))=1$.
\end{definition}

An important result from \cite{ACR98} is the following:
\begin{corollary}[Corollary 3.2 in \cite{ACR98}]
\label{corollary8.4}
  Let
  $$
  k(n)=O(\log n).
  $$
  If there exists a quick HSG
  $$
  H: k(n)\rightarrow n,
  $$
  then
  $$
  \mathcal{BPP}=\mathcal{P}.
  $$
\end{corollary}

We are now ready to prove Theorem \ref{theorem1.11}:

\vskip 0.2 cm
\noindent{\em Proof of Theorem \ref{theorem1.11}.} Let 
$$
k(n)=O(\log n).
$$
Suppose, for the sake of contradiction, that there exists a quick HSG 
$$
H: k(n)\rightarrow n.
$$ 
Then, by Corollary \ref{corollary8.4}, we would have 
$$
\mathcal{P}=\mathcal{BPP},
$$
which contradicts Theorem \ref{theorem1.1}. This completes the proof.\Q.E.D

\vskip 0.3 cm
\section{Concluding Remarks and Open Problems}
\label{sec:conclusion}
\vskip 0.3 cm

In conclusion, we have shown that there exists a language $L_d$ that is accepted by some probabilistic Turing machine with bounded error probability $1/3$, but not by any polynomial-time deterministic Turing machine. To achieve this, we first encode any single-tape polynomial-time deterministic Turing machine into an integer using the method presented in \cite{AHU74}. We then encode the order (i.e., the polynomial degree) of the machine into a binary string. By concatenating these two binary strings, we establish a $(1,1)$ correspondence $e$ between $\mathbb{N}_1$ and the set of all pairs $\{(M,k)\}$ of polynomial-time deterministic Turing machines and their orders, where any integer that cannot be decoded is treated as representing the trivial polynomial-time deterministic Turing machine. These steps follow the approach of \cite{Lin21a}.

Next, we design a four-tape probabilistic Turing machine $M_0$ that diagonalizes against all polynomial-time deterministic Turing machines. Theorem \ref{theorem4.1} describes in detail the operation of $M_0$ and shows that there exists a language $L_d$ accepted by $M_0$ but by no polynomial-time deterministic Turing machine. In Theorem \ref{theorem4.2}, we analyze the running time of $M_0$ and prove that it runs within time $$O(n^k)$$ for every $k\in\mathbb{N}_1$. We further show in Theorem \ref{theorem5.1} that $$L_d\in\mathcal{BPP}. $$ It thus follows that the Theorem \ref{theorem1.1} holds.

By slightly modifying the proof of Theorem \ref{theorem1.1}, we prove Theorem \ref{theorem1.3} in Section \ref{sec:proof_of_pneqrp} and Theorem \ref{theorem1.5} in Section \ref{sec:proof_of_pneqcorp}, establishing that 

$$\mathcal{P}\subsetneqq\mathcal{RP}$$ and $$\mathcal{P}\subsetneqq{\rm co}\mathcal{RP}. $$ Similar arguments also yield $$\mathcal{P}\subsetneqq\mathcal{ZPP}. $$

Our result in Theorem \ref{theorem1.1} disproves the conjecture that $$\mathcal{P}=\mathcal{BPP}. $$ Furthermore, in Section \ref{sec:randomness_helps} we prove that randomness is essential and useful in probabilistic algorithm design. Specifically, we show that the number of random bits used by any probabilistic algorithm accepting the language $L_d$ cannot be reduced to $O(\log n)$.

We also provide negative answers regarding the existence of certain efficient {\em pseudorandom generators} (PRGs) and efficient {\it quick hitting set generators} (HSGs). These results are summarized in Theorem \ref{theorem1.10} and Theorem \ref{theorem1.11} in Section \ref{sec:randomness_helps}.

Moreover, the question of whether {\it quantum computers} are strictly more powerful than {\em classical probabilistic computers} remains open. In particular, proving $$\mathcal{BPP}\subsetneqq\mathcal{BQP}$$ would constitute a major breakthrough in complexity theory (see e.g. \cite{NC00}). One reason is that we do not currently know how to simulate an arbitrary probabilistic Turing machine on a quantum Turing machine while correctly preserving the acceptance probability (i.e., how to ``flip the answer" when necessary). In addition, while the complexity class $\mathcal{NP}$ is known to have complete problems (see e.g., \cite{Coo71}), it is unclear whether $\mathcal{BPP}$ has a comparably rich structure. Although we have shown that $$L_d\not\in\mathcal{P}$$ but $$L_d\in\mathcal{BPP},$$ we do not know whether the language $L_d$ is a $\mathcal{BPP}$-intermediate language (in the sense of Ladner \cite{Lad75}, who showed the existence of $\mathcal{NP}$-intermediate languages assuming $\mathcal{P}\ne\mathcal{NP}$).

For the probabilistic Turing machine model (Definition \ref{definition2.8}),it is unknow whether $${\rm BPTIME}[T(n)]\subsetneqq{\rm BPTIME}[t(n)] $$ holds for time constructible functions $T(n)$ and $t(n)$ satisfying $$\lim_{n\rightarrow\infty}\frac{T(n)\log T(n)}{t(n)}=0.$$ The pioneering works on this fundamental problem include  \cite{Bar02,FS04}.

Finally, the analogous question for the quantum Turing machine model (Definition \ref{definition2.4}) is also open: it is unknown whether $${\rm BQTIME}[T(n)]\subsetneqq{\rm BQTIME}[t(n)]$$ holds for time-constructible functions $T(n)$ and $t(n)$ satisfying  $$\lim_{n\rightarrow\infty}\frac{T(n)\log T(n)}{t(n)}=0.$$

\bibliographystyle{ACM-Reference-Format}

%\clearpage
\appendix
\section{Appendix: Two Definitions of Probabilistic Turing Machines Are Equivalent}
\label{sec:appendix}

We show that Definition \ref{definition2.8} and Definition \ref{definition2.ten} are equivalent.

We use the following notation: Let $M$ be a probabilistic Turing machine. Then $L(M)$ denotes the language accepted by $M$. We first prove the ``only if" direction, i.e., Definition \ref{definition2.8} $\Longrightarrow$ Definition \ref{definition2.ten}.

\begin{lemma}
Let $M_1$ be a probabilistic Turing machine of the probabilistic transition type (Definition \ref{definition2.8}). Then there exists a random-tape probabilistic Turing machine (Definition \ref{definition2.ten}) $M_2$ that simulates $M_1$, i.e., $L(M_1)=L(M_2)$.
\end{lemma}
\begin{proof}
For each probabilistic transition in $M_1$, suppose there are $n$ possible outcomes with probabilities $p_1$, $p_2$, $\cdots$, $p_n$, where $p_i=\frac{a_i}{m}$ for integers $a_i$ and $m$ such that $\sum\limits_{i=1}^na_i=m$.

Let $k$ be the smallest integer such that $m\leq 2^k$. We represent each probability $p_i$ in the form $\frac{a_i'}{2^k}$ by setting $a'_i=\lfloor p_i\cdot 2^k\rfloor$ and then adjust the $a'_i$ minimally (if necessary) so that $\sum\limits_{i=1}^na'_i=2^k$.

We assume the $p_i$ are rational (so that such a finite $k$ exists). This yields an exact representation for $\mathcal{PP}$-class machines and a negligible small error for $\mathcal{BPP}$-class machines.

We now construct a random-tape probabilistic Turing machine $M_2$ that simulates $M_1$. At each step where $M_1$ makes a probabilistic transition:
\begin{itemize}
  \item {$M_2$ reads $k$ bits from the random tape and interprets them as an integer $b\in\{0,1,\cdots,2^k-1\}$.}
  \item {It partitions the interval $[0,2^k)$ into $n$ subintervals $I_1$, $I_2$, $\cdots$, $I_n$ of lengths $|I_i|=a_i'$.}
  \item {It execute the $i$-th transition if $b\in I_i$.}
\end{itemize}

The probability that $b\in I_i$ is exactly $\frac{a_i'}{2^k}=p_i$ (or exactly $p_i$ when no adjustment was needed). Since the random-tape bits are independent and uniformly distributed, there is no correlation between successive simulation steps.

Each simulation step requires $O(k)$ time. When $k=O(\log n)$, the overhead is polynomial for $\mathcal{BPP}$ and $\mathcal{PP}$) computations. Therefore, $L(M_1)=L(M_2)$.
\end{proof}

\vskip 0.5cm

We now prove the ``if" direction, i.e., Definition \ref{definition2.ten} $\Longrightarrow$ Definition \ref{definition2.8}.

\begin{lemma}
Let $M_2$ be a random-tape probabilistic Turing machine (Definition \ref{definition2.ten}). Then there exists a probabilistic Turing machine $M_1$ of the probabilistic transition type (Definition \ref{definition2.8}) that simulates $M_2$, i.e., $L(M_1)=L(M_2)$.
\end{lemma}

\begin{proof}
Let $M_2$ be a random-tape probabilistic Turing machine. At any given step, $M_2$ may read up to $k$ bits from the random tape, yielding $2^k$ possible outcomes, each with probability $\frac{1}{2^k}$.

We construct a probabilistic Turing machine $M_1$ that simulates $M_2$ as follows. $M_1$ enumerates all $2^k$ bit sequences of length $k$. For each possible sequence $s\in\{0,1\}^k$, $M_1$ includes a transition of probability $\frac{1}{2^k}$ that replicates the behavior of $M_2$ upon reading $s$.

If $M_2$ reads bits dynamically (e.g., reading one bit and then conditionally reading more), $M_1$ uses extended states to track the history of simulated random bits. Specifically, the states of $M_1$ are augmented with a ``random prefix" component (e.g., of the form $q\times\{0,1\}^*$) that records the bits simulates so far.

Each bit read by $M_2$ is simulated in $M_1$ by a binary probabilistic branch of probability $\frac{1}{2}$. Cumulative probabilities (such as $\frac{1}{4}$ for a specific pair of bits) are thus encoded exactly. The extended states ensure that dependencies between bits are respected under dynamic reading. 

Since any halting computation of $M_2$ on an input of length $n$ uses only finitely many random bits, $M_1$ generates precisely the required bits and avoids unnecessary infinite branching.

Each simulation step incurs an $O(k)$ overhead. When $k$ is polynomial in the input size, the total simulation overhead remains polynomial. Therefore, $L(M_1)=L(M_2)$.

\end{proof}
\end{document}